\documentclass[twocolumn,  tighten,  twocolappendix]{aastex631} %linenumbers

\shorttitle{Impact of Galaxy Mergers on Early-type Galaxies}
\shortauthors{Yoon et al. (2023)}

\begin{document}

\title{Impact of Galaxy Mergers on Stellar Population Profiles of Early-type Galaxies}

\email{yyoon@kasi.re.kr}

\author[0000-0003-0134-8968]{Yongmin Yoon}
\affiliation{Korea Astronomy and Space Science Institute (KASI), 776 Daedeokdae-ro, Yuseong-gu, Daejeon, 34055, Republic of Korea}

\author[0000-0002-9434-5936]{Jongwan Ko}
\affiliation{Korea Astronomy and Space Science Institute (KASI), 776 Daedeokdae-ro, Yuseong-gu, Daejeon, 34055, Republic of Korea}
\affiliation{University of Science and Technology, Gajeong-ro, Daejeon 34113, Republic of Korea}

\author[0000-0002-1710-4442]{Jae-Woo Kim}
\affiliation{Korea Astronomy and Space Science Institute (KASI), 776 Daedeokdae-ro, Yuseong-gu, Daejeon, 34055, Republic of Korea}

\begin{abstract}
We study the impact of galaxy mergers on stellar population profiles/gradients of early-type galaxies (ETGs) using ETGs at $z<0.055$ in the Stripe 82 region of the Sloan Digital Sky Survey and MaNGA integral field unit spectroscopic data. Tidal features around ETGs, which are detected from deep coadded images, are regarded as direct observational evidence for recent mergers. We find that ETGs with tidal features have less negative metallicity gradients and more positive age gradients than ETGs without tidal features at $M_\mathrm{star}\gtrsim10^{10.6}M_\odot$. Moreover, when integrating all the resolved stellar populations, ETGs with tidal features have lower metallicities by $\sim0.07$ dex and younger ages by $\sim1$ -- $2$ Gyr than ETGs without tidal features. Analyzing star formation histories, we discover that the mass fraction of young stellar populations with age $<5$ Gyr is higher in the central regions of ETGs with tidal features than in the same regions of the counterparts without tidal features. Compared to normal ETGs, ETGs with tidal features have a slow metal-enrichment history in the early universe, but they have been accelerating the metal enrichment through recently formed stars over the last few billion years. Many of our results can be explained if the effects of recently occurred mergers are different from those in the early universe, which are more likely to be rich in gas.
\end{abstract}
\keywords{Early-type galaxies (429) --- Galaxy ages (576) --- Galaxy evolution (594) --- Galaxy mergers (608) --- Galaxy stellar content (621) --- Tidal tails (1701)}

\section{Introduction}\label{sec:intro}

It is known that early-type galaxies (ETGs) already formed and accreted a large fraction of currently existing stars a long time ago. Through these processes, they consumed available cold gas, which is a source of star formation activity. Thus, they are evolved systems with a low level of star formation activity and old stellar populations (age $\gtrsim5$ Gyr) whose color is red ($g-r\gtrsim0.7$) in the optical bands \citep{Park2005,Gallazzi2006,Graves2009,Choi2010,Schawinski2014,Gonzalez2015,Li2018,Lacerna2020,YP2020}. Stars in ETGs are more concentrated in the central regions of galaxies than late-type galaxies \citep{Park2005,Choi2010}, so that their surface brightness profiles can be described by centrally concentrated light profiles, such as the de Vaucouleurs profile or the S{\'e}rsic profiles, with high S{\'e}rsic indices of $n\gtrsim3$ \citep{Blanton2009,Huertas-Company2013}. Another general photometric property of ETGs is that they have relatively smooth and simple morphologies compared to spiral galaxies, which exhibit two clearly distinguishable components (red bulges and blue disks) and complex features such as spiral arms and sometimes blobs of highly star-forming regions \citep{Nair2010}.

Spatially resolved kinematic properties for a large number of ETGs have been revealed through recent large surveys based on integral field unit (IFU) spectroscopy \citep{Bacon2001,Bershady2010,Cappellari2011,Sanchez2012,Sanchez2016,Ma2014,Croom2021,Yoon2021}. The large survey data based on IFU observation have shown new aspects of kinematics of ETGs. For instance, it has been found that ETGs are not simply pressure-supported systems dominated by random orbits of stars, but a large fraction of them have substantial rotating stellar components that are analogous to rotating disks in late-type galaxies \citep{Cappellari2016,Graham2018}, even though they have round and nondisky shapes. Therefore, many studies used a new classification scheme dividing ETGs into slow or fast rotators based on resolved kinematics \citep{Emsellem2007,Emsellem2011,Jesseit2009,Khochfar2011,Cappellari2016,Graham2018}. This is increasingly being considered a more physically meaningful classification than the traditional separation into ellipticals and lenticulars, which is heavily influenced by line-of-sight inclinations \citep{Emsellem2007,Cappellari2011,Cappellari2016}.

Radial structures of stellar population properties for diverse ETGs have also been investigated through spectroscopic observations and IFU survey data. These studies generally suggested that ETGs have negative metallicity profiles and mildly positive or flat age profiles, indicating that stars in the inner parts of ETGs have higher metallicities and slightly younger ages than those in the outer regions \citep{Koleva2011,Goddard2017,SanRoman2018,Lacerna2020,Neumann2021}.   

A galaxy merger is one of the most crucial factors in the formation and growth of ETGs in the $\Lambda$ cold dark matter ($\Lambda$CDM) universe \citep{Baugh1996,Christlein2004,DeLucia2006,DeLucia2007,Wilman2013}, particularly for massive galaxies \citep{Yoon2017,YP2022}. 
Mergers of galaxies with abundant gas cause a significant increase in star formation activity \citep{Hernquist1989,Mihos1996,Springel2005}, which leads to the rapid exhaustion of available cold gas. In the end, merger remnants become red and quiescent within a time shorter than $\sim1$ Gyr \citep{Springel2005,Hopkins2008b,Brennan2015}. Strong energy output of feedback effects from active galactic nuclei (AGNs) triggered during merger processes can also play a role in quenching post-merger galaxies \citep{Hopkins2005,Hopkins2008b,Springel2005}. Galaxy mergers are able to significantly affect stellar kinematics such as angular momentum \citep{Cox2006,Robertson2006,Emsellem2007,Hoffman2010,Lagos2017,Penoyre2017,Yoon2022}. Furthermore, galaxy mergers can cause concentrated stellar light distributions in post-merger galaxies \citep{Barnes1988,Naab2006,Hilz2013}, while dry minor mergers are responsible for the substantial size growth of ETGs \citep{Bernardi2011a,Oogi2013,Yoon2017}.

Formation and growth histories of ETGs and physical mechanisms involved in the evolution, especially galaxy mergers, are imprinted on the radial stellar population profiles of ETGs, such as metallicity and age profiles. Thus, many previous studies conducted simulations to determine how galaxy mergers affect radial stellar population profiles/gradients. As a simple example, several simulation studies showed that gas-rich mergers can cause negatively steep metallicity gradients and positive age gradients \citep{Kobayashi2004,Hopkins2009,Cook2016,Taylor2017} through inflow of gas into the centers and subsequent central high star formation activity \citep{Hernquist1989,Barnes1991,Barnes1996,Hopkins2008a,Hopkins2008c}. On the other hand, dry mergers are capable of flattening metallicity profiles by mixing stars \citep{White1980,Kobayashi2004,DiMatteo2009,Taylor2017}.

Since we do not observe the stream of cosmic-scale time, but only a snapshot of the universe, a direct determination of the effects of mergers on the properties of ETGs through observational data is not as easy as simulation works. However, galaxy mergers leave traces of gravitational interactions called tidal features in the shape of tails, streams, and shells \citep{Toomre1972,Quinn1984, Barnes1988,Hernquist1992,Feldmann2008}. Hence, tidal features around ETGs are considered the most direct observational evidence for recent mergers. Since the surface brightness of tidal features is typically lower than that of galaxy main bodies, it requires deep images to detect and study tidal features. 

Using the fact that tidal features are direct evidence for recent mergers, several previous observational studies examined the impact of galaxy mergers on photometric \citep{Schweizer1992,Tal2009,Schawinski2010,Kaviraj2011,Hong2015,YL2020} and kinematic \citep{Krajnovic2011,Duc2015,Oh2016,Yoon2022} properties of ETGs. For example, it is found that ETGs with blue optical color are more likely to have tidal features or morphological disturbances \citep{Schweizer1992,Tal2009,Schawinski2010,Kaviraj2011}, implying that young stellar populations in ETGs can be related to recent mergers. \citet{Hong2015} discovered that nearly half of the luminous AGN hosts (mostly ETGs) have tidal features unlike normal ETGs, which have a lower fraction, suggesting that luminous AGNs in ETGs are likely to be triggered by recent mergers. \citet{YL2020} investigated how the fraction of ETGs with tidal features depends on the age and internal structure of ETGs. By doing so, they found that young ETGs with centrally concentrated light profiles and ETGs with dust lanes are likely to have experienced recent galaxy mergers. Combining tidal features detected in deep images with IFU spectroscopic data, \citet{Yoon2022} suggested that galaxy mergers can affect the stellar kinematics of ETGs by increasing or reducing stellar angular momentum, depending on the gas abundance in merger processes. Moreover, \citet{Yoon2022} showed that kinematic misalignments\footnote{Misalignment of the photometric position angle from the galaxy light distribution and the kinematic position angle derived from the observed stellar velocity map.} in ETGs can also be influenced by galaxy mergers.

Here, we investigate how galaxy mergers affect stellar population (metallicity and age) profiles/gradients of ETGs using ETGs in the Stripe 82 region of the Sloan Digital Sky Survey (SDSS) with Mapping Nearby Galaxies at Apache Point Observatory (MaNGA; \citealt{Bundy2015,Drory2015,Yan2016,Wake2017}) IFU spectroscopy data. Radial stellar population profiles of ETGs from MaNGA data combined with tidal features detected from deep coadded images of the Stripe 82 enable us to constrain the formation/evolution histories of ETGs as well as the role of galaxy mergers on the stellar populations in ETGs.

In this paper, we use $H_0=67.8$ km s$^{-1}$ Mpc$^{-1}$, and $\Omega_m=0.308$ with a flat $\Lambda$CDM cosmology \citep{Plank2016}.
\\

\section{Data and Analysis}\label{sec:data}

\subsection{SDSS-IV MaNGA}\label{sec:manga}
The MaNGA IFU spectroscopic survey, which is a project of the fourth generation of the SDSS \citep{Blanton2017}, obtained observation data using the ARC 2.5 m telescope \citep{Gunn2006}. The observation was carried out with 17 fiber-bundled IFUs with sizes of $12\arcsec$ -- $32\arcsec$, depending on the number of fibers. The 17 IFUs are assigned within the $3\degr$ field of view. The spectrograph of the MaNGA IFU survey project is the same as that of the Baryonic Oscillation Spectroscopic Survey \citep{Smee2013}, whose wavelength coverage is 3600 -- 10300\,\AA, and the medium-range spectral resolution is $R\sim2000$. The target galaxies of MaNGA project were adopted according to criteria for $i$-band absolute magnitude and redshift (and near-ultraviolet $-\,i$ color for a small number of galaxies), ensuring that targets are evenly distributed in color--magnitude space and have a uniform spectroscopic coverage up to the 1.5 or 2.5 half-light radius along the major axis ($R_e$).\footnote{We note that $R_e$ was derived based on the $r$-band elliptical Petrosian flux.} We refer to \citet{Wake2017} for more details about the selection of the target galaxies.
\\

\begin{figure*}
\includegraphics[scale=0.35]{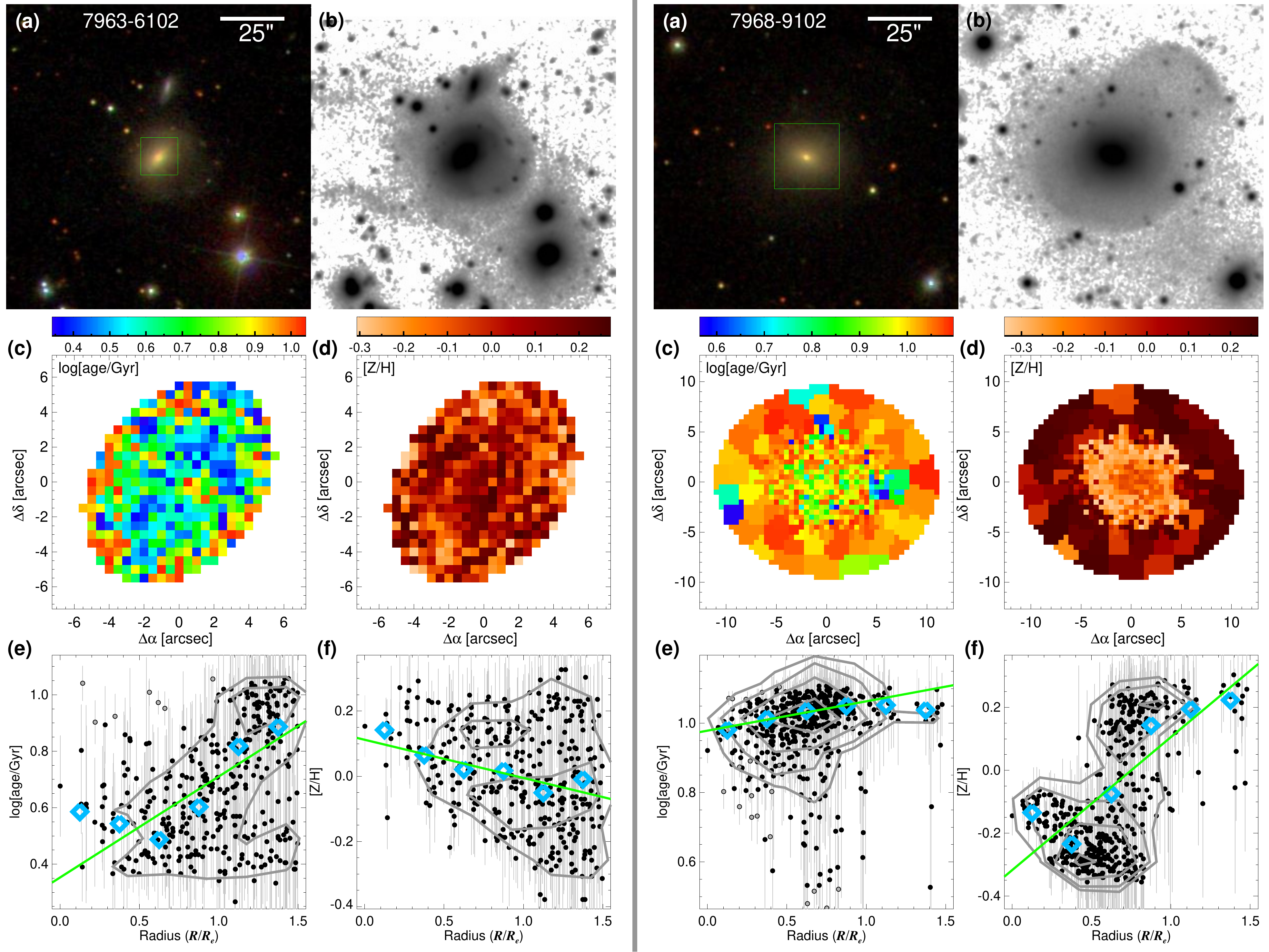}  
\centering
\caption{Examples of ETGs with tidal features. Panel (a): single-epoch color images of SDSS. The horizontal bar in the color image indicates the angular scale of the image. The plate ID (e.g., 7963) and IFU design ID (e.g., 6102) are displayed in the color image. The green square denotes the window size of the two-dimensional age/metallicity maps shown in panels (c) and (d). Panel (b): deep coadded images of Stripe 82. The angular scale of the deep image is identical to that of the color image. Panels (c) and (d): two-dimensional maps for the spatially resolved age and metallicity within $1.5R_e$. Color bars above the panels indicate color-coded age and metallicity. Panels (e) and (f): radial age and metallicity profiles. The black dots with the gray error bars indicate ages/metallicities of Voronoi cells with their uncertainties. The blue diamonds are the error-weighted mean values of ages/metallicities within the bins (bin size: $0.25R_e$). The green line denotes the linear function fitted to ages/metallicities with their errors after excluding outliers. The gray dots are the outliers excluded in the line fittings. The slope of the fitted line is the age/metallicity gradient. The level of gray contours indicates the densities of the data points.
\label{fig:ex_1}}
\end{figure*} 

\begin{figure*}
\includegraphics[scale=0.35]{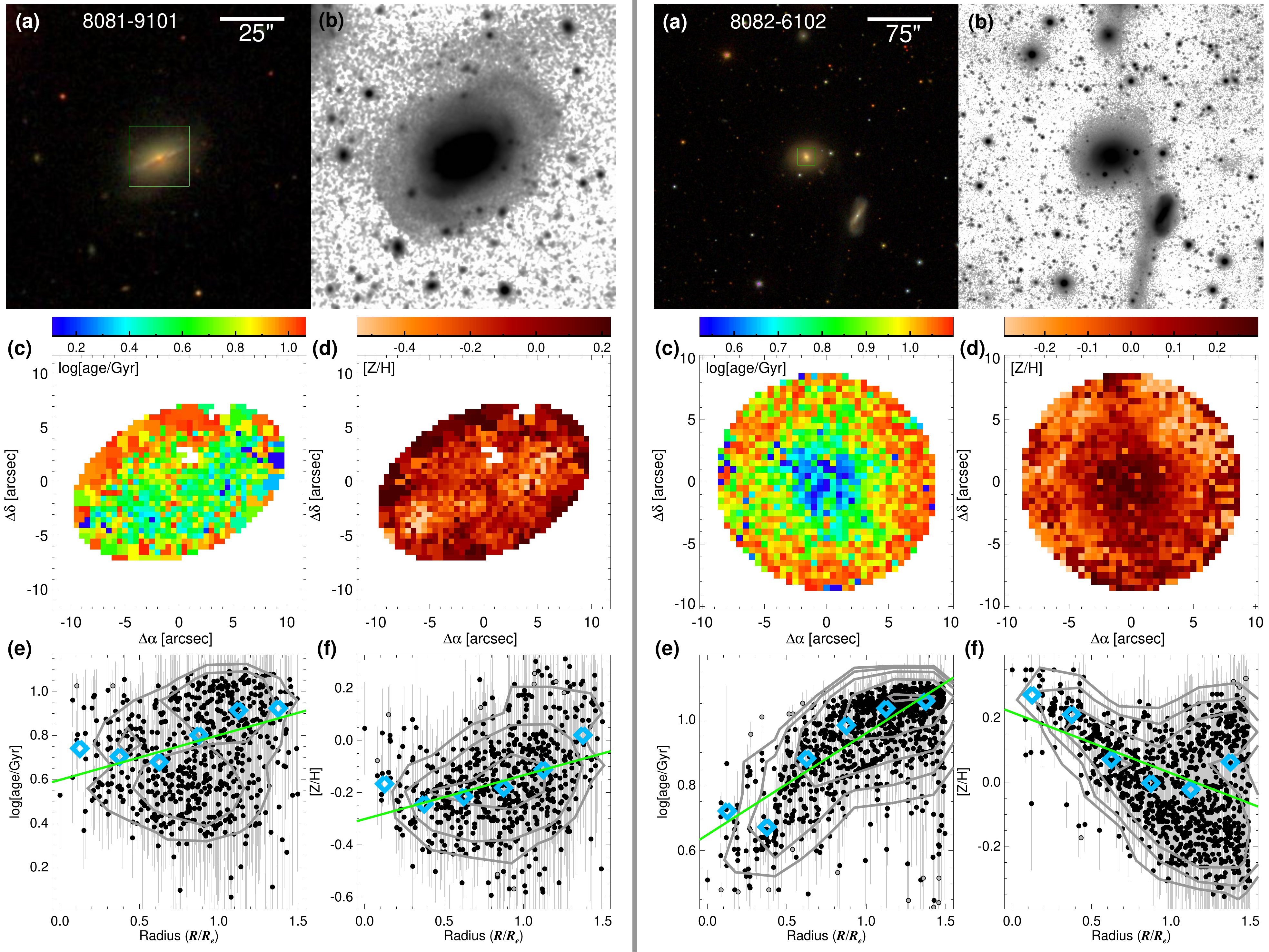}  
\centering
\caption{Examples of ETGs with tidal features. Descriptions about the figure are identical to those in Figure \ref{fig:ex_1}. 
\label{fig:ex_2}}
\end{figure*} 

\begin{figure*}
\includegraphics[scale=0.35]{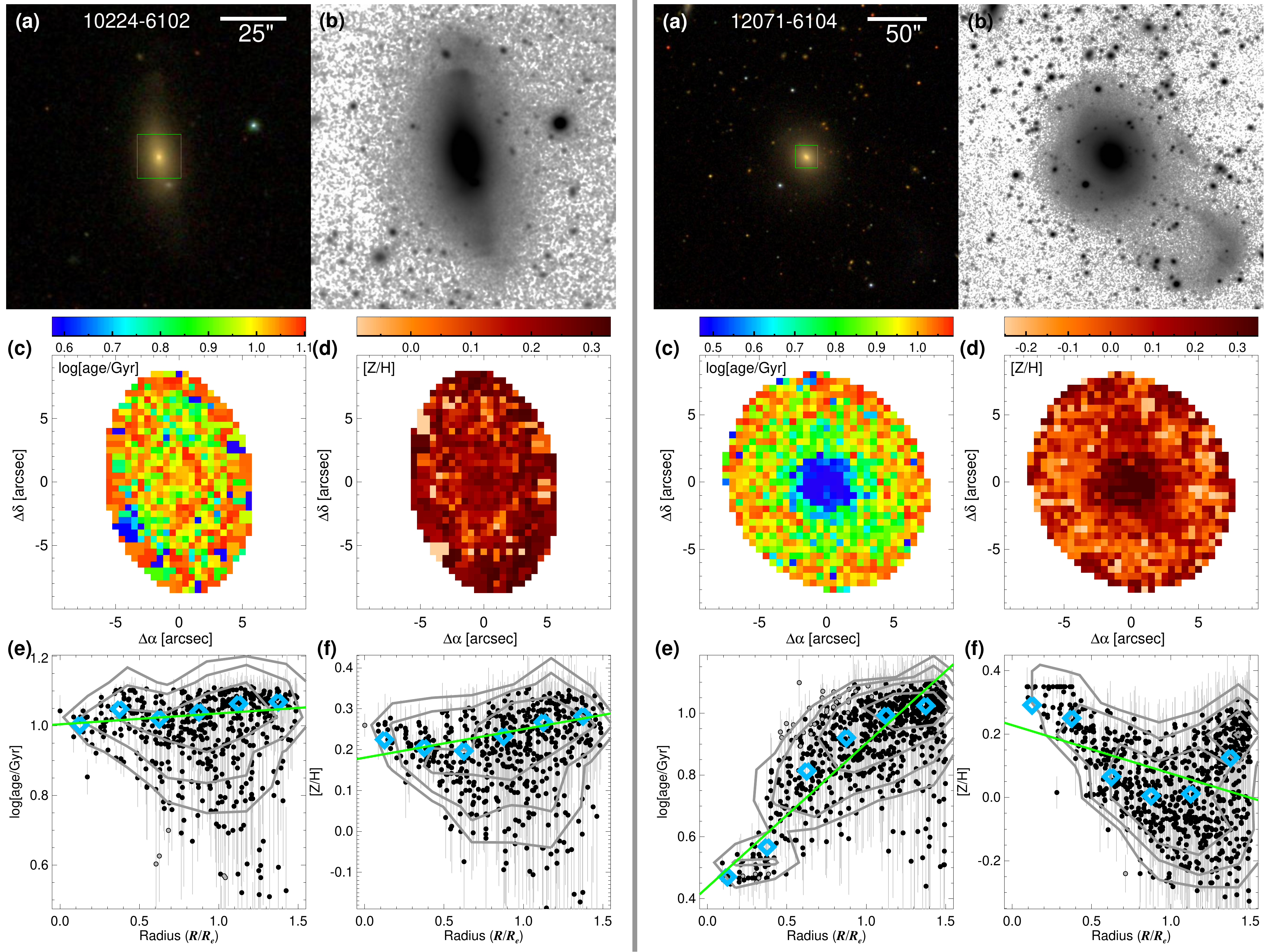}  
\centering
\caption{Examples of ETGs with tidal features. Descriptions about the figure are identical to those in Figure \ref{fig:ex_1}. 
\label{fig:ex_3}}
\end{figure*} 

\begin{figure*}
\includegraphics[scale=0.35]{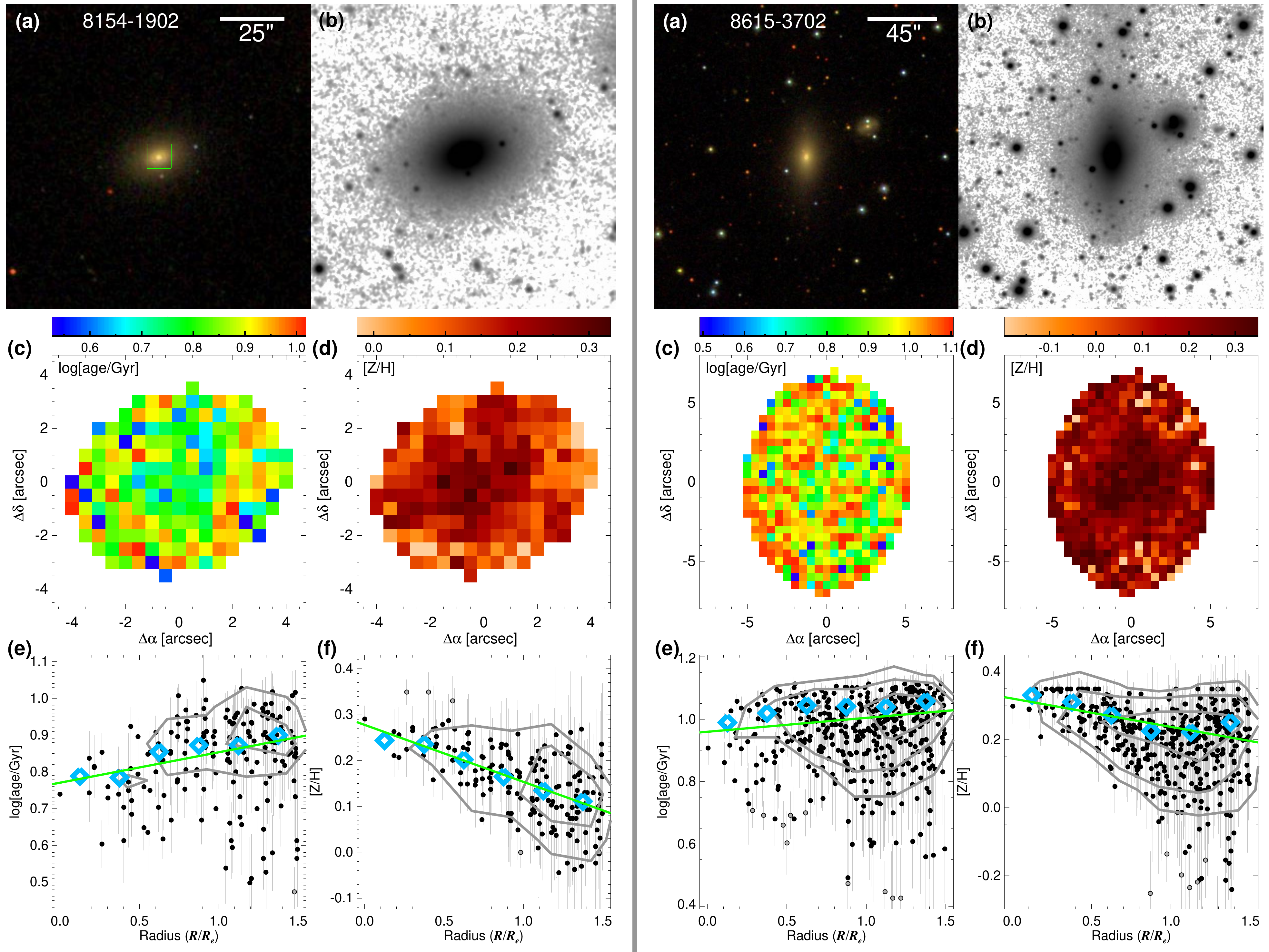}  
\centering
\caption{Examples of normal ETGs without tidal features. Descriptions about the figure are identical to those in Figure \ref{fig:ex_1}.
\label{fig:ex_4}}
\end{figure*} 

\begin{figure*}
\includegraphics[scale=0.35]{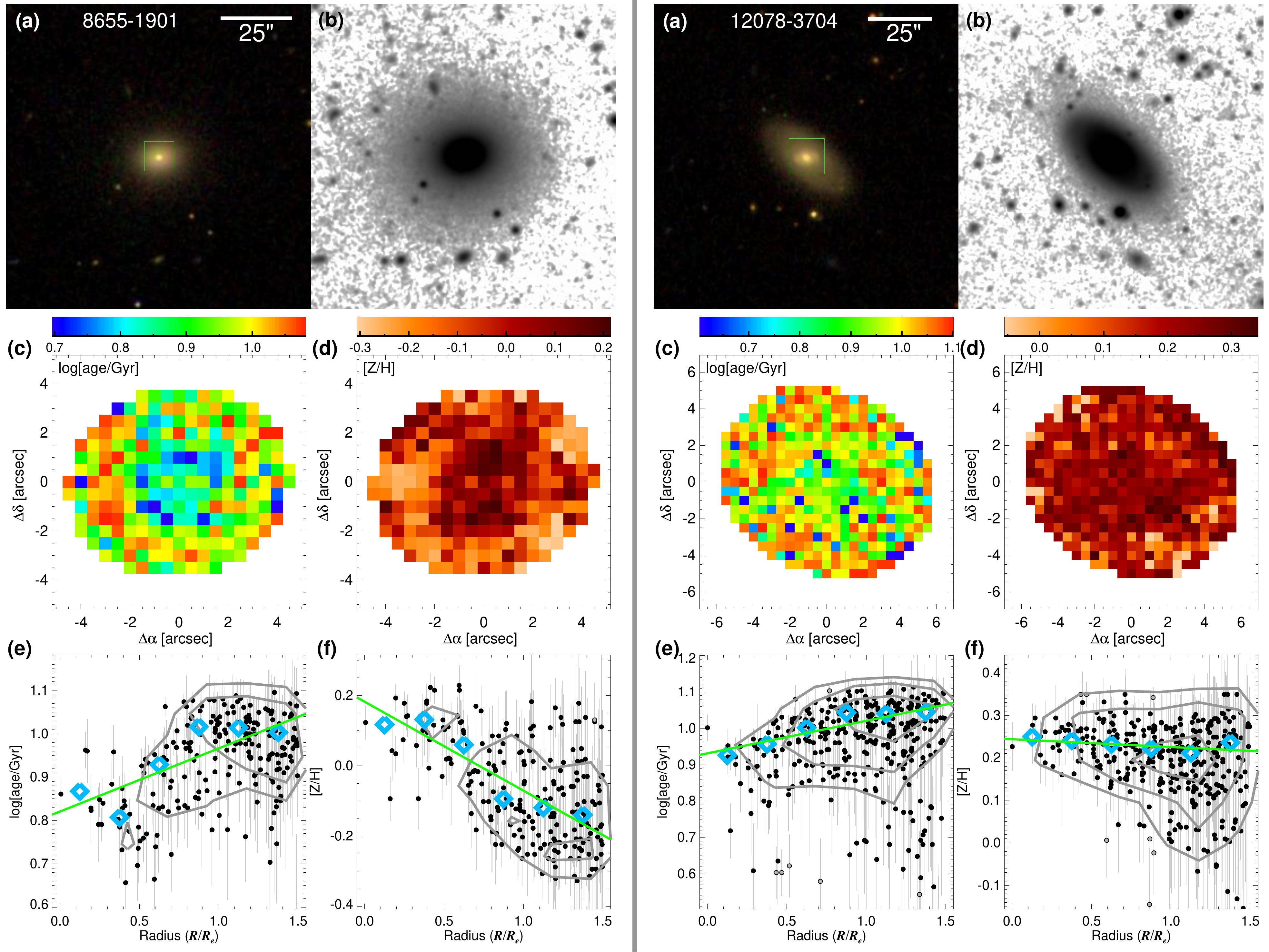}  
\centering
\caption{Examples of normal ETGs without tidal features. Descriptions about the figure are identical to those in Figure \ref{fig:ex_1}.
\label{fig:ex_5}}
\end{figure*} 

\subsection{Resolved Stellar Population Properties}\label{sec:rspp}
In this study, spatially resolved stellar population properties for galaxies in the MaNGA survey are from the MaNGA FIREFLY value-added catalog (VAC; \citealt{Neumann2022}), which contains information for 10,010 unique MaNGA galaxies in the final release version (SDSS data release 17). The catalog provides spatially resolved stellar population information for each galaxy, such as ages, metallicities, stellar masses, dust attenuations, and star formation histories. These stellar population properties were derived from the full spectrum fitting code called fitting iteratively for likelihood analysis (FIREFLY; \citealt{Wilkinson2017}). FIREFLY is a $\chi^2$-minimization fitting code that iteratively fits a input spectral energy distribution to determine the best-fit convergence using the Bayesian information criterion and combinations of single stellar population (SSP) models. To construct the catalog, FIREFLY was conducted on MaNGA IFU data cubes that were processed through the data analysis pipeline (DAP; \citealt{Belfiore2019,Westfall2019}). The DAP includes adaptive Voronoi spatial binning \citep{Cappellari2003} to ensure a minimum signal-to-noise ratio (S/N) $\sim10$ for the stellar continuum of each spatial bin. Using the full spectrum fitting code called penalized pixel-fitting (pPXF; \citealt{Cappellari2004,Cappellari2017}), the DAP determines stellar kinematics (velocities and velocity dispersions) and emission line spectra, which are input to the FIREFLY code.

The FIREFLY VAC used the dust law in \citet{Calzetti2000} and SSP models that are based on the Kroupa initial mass function \citep{Kroupa2001}. The catalog provides two structurally identical result sets that differ only in the stellar population model library used to fit the input spectra. The first set uses the M11-MILES model templates from \citet{Maraston2011}, which are based on the MILES stellar library \citep{SanchezBlazquez2006}. The wavelength coverage of the M11-MILES models reaches from 3500 to 7430\,\AA. The metallicity grid spans $-2.25\le[Z/H]\le0.35$ with 10 steps, while the age grid covers from 0.0065 to 15.0 Gyr with 50 steps. The second set uses the MaStar SSP models described in \citet{Maraston2020}, which are based on the MaStar stellar library \citep{Yan2019}. The models cover the wavelength range of 3600 -- 10,300\,\AA, with metallicities covering $-2.25\le[Z/H]\le0.35$ and ages spanning from 0.003 to 15.0 Gyr.

Although the wavelength coverage of the MaStar models is wider than that of the M11-MILES models, we use the M11-MILES model throughout this study for two reasons. The first reason is that the metallicity enrichment history for stellar populations of ETGs is more realistic when using the M11-MILES models than the MaStar models, as shown in Figure \ref{fig:spcomp} in Appendix \ref{Appendix_A}. In the figure, the M11-MILES model shows that more recently formed stars in ETGs have higher metallicities than stars that formed in the early universe. However, the MaStar model displays the opposite trend that more recently formed stars in ETGs have lower metallicities. Therefore, the metal-enrichment history from the M11-MILES is consistent with the general history of cosmic metal enrichment \citep{Edvardsson1993,Fynbo2006,Maiolino2008,Mannucci2009,Piatti2013,Zahid2013,Zahid2014,Madau2014,Guo2016,Valentini 2018}, whereas that from the MaStar model contradicts it. The second reason is that stellar population models based on the MILES stellar library have been widely used in previous studies of global or resolved stellar populations of galaxies (e.g., \citealt{Koleva2011,Goddard2017,Zheng2017,Li2018,Lian2018,SanRoman2018,Ferreras2019,Oyarzun2019,Lacerna2020,Zibetti2020,Neumann2021}). So, the use of the M11-MILES model facilitates comparisons with other studies. Although we use the M11-MILES models in this study, results for stellar population gradients from the MaStar models, which are described in Appendix \ref{Appendix_A}, are roughly consistent with those from M11-MILES at several points. See \citet{Neumann2022} for an intensive comparison between the M11-MILES and MaStar models.

The FIREFLY code gives a full star formation history with light weights and mass weights of the individual SSPs for an input spectrum. 
In this study, we use mass-weighted ages or metalicities that are linearly averaged over the SSPs,
\begin{equation}
P_\mathrm{MW}=\sum_{i=1}^{n} w_iP_i.
\label{eq:par1}
\end{equation}
In this equation, $P$ is a stellar population parameter such as age or metallicity, and $w_i$ is the mass weight of the $i$th SSP. Stellar population parameters within a certain ellipse or annulus are calculated as follows:
\begin{equation}
P_\mathrm{area}=\frac{\sum_{i=1}^{n}  m_i P_{\mathrm{MW},i}}{\sum_{i=1}^{n} m_i},
\label{eq:par2}
\end{equation}
where $n$ is the total number of spaxels within the area, $m_i$ is the stellar mass of the $i$th spaxel, and  $P_{\mathrm{MW},i}$ is the mass-weighted age or metallicity of the $i$th spaxel calculated by Equation (\ref{eq:par1}).\footnote{Total metallicity and age for a galaxy are calculated using Equation (\ref{eq:par2}) and the stellar population information of all spaxels within $1.5R_e$.} Mass-weighted stellar population parameters have the advantage that they explore stellar populations with diverse ages, as opposed to light-weighted ones that are very sensitive to recently formed stars. We note that the results for stellar population gradients from light-weighted parameters, which are shown in Appendix \ref{Appendix_B}, are in agreement with those from mass-weighted ones. 

The spatial coverage for MaNGA IFU data reaches up to $1.5$ or $2.5R_e$ of galaxies (mainly 1.5$R_e$ for galaxies at $z<0.055$, which is the upper redshift limit for our final sample). In this study, we only used spaxels within 1.5$R_e$ of galaxies to consistently examine stellar population profiles of all the ETGs in the final sample.

To derive gradients for stellar population profiles or to fit linear functions to stellar population properties versus stellar mass, we used the line-fitting code LTS\_LINEFIT\footnote{\url{https://www-astro.physics.ox.ac.uk/~mxc/software/}} \citep{Cappellari2013}, which performs the $\chi^2$ minimization after trimming outliers. This fitting code uses the least trimmed squares robust technique that clips data points from inside out \citep{Cappellari2013}, which is opposite to the standard clipping method. This method means that the fitting result is less strongly affected by catastrophic outliers. 

We show examples of two-dimensional maps for the spatially resolved age and metallicity for several ETGs in Figures \ref{fig:ex_1}, \ref{fig:ex_2}, \ref{fig:ex_3}, \ref{fig:ex_4}, and \ref{fig:ex_5}. In these figures, we also display examples of radial age and metallicity profiles. Throughout this paper, the age/metallicity gradient ($\nabla[Z/H]$/$\nabla(\log\mathrm{age})$) is defined as the slope of the linear function fitted to radial profiles of $\log(\mathrm{age/Gyr})$ or $[Z/H]$ of Voronoi cells,\footnote{The errors of ages and metallicities were also taken into account in the fitting process.} excluding $3\sigma$ outliers.\footnote{On average, $\sim1$--$3\%$ of Voronoi cells in a galaxy in our sample are trimmed in the line fitting to the radial profiles. Less than $9.2\%$ of the cells are excluded by the fitting algorithm in $94\%$ of galaxies in our sample. Thus, excluding outliers in the fitting procedure does not significantly affect the derivation of stellar population gradients.} The unit of stellar population gradients is dex/$R_e$. 
\\

\subsection{ETG Sample}\label{sec:sample}

The Stripe 82 region includes 699 MaNGA galaxies in the coadded images, 320 of which are ETGs.\footnote{In the final  sample, we excluded one ETG hosting a type 1 quasar.} ETGs were classified from the sample by visually inspecting combined color images of the $g$, $r$, and $i$ bands in SDSS (e.g., panels (a) in Figures \ref{fig:ex_1}, \ref{fig:ex_2}, \ref{fig:ex_3}, \ref{fig:ex_4}, and \ref{fig:ex_5}) in the same way as in \citet{Yoon2021} (and also \citealt{Yoon2022}), in which we visually classified the galaxy morphology for more than 2000 MaNGA galaxies. Our morphology classifications are well consistent with other classification results from \citet{Dominguez2018} and \citet{Vazquez2022}\footnote{Specifically, our $T$-type morphology classifications are consistent with those from the two results \citep{Dominguez2018,Vazquez2022} within about $\pm1$.} (see \citealt{Yoon2021}). ETGs in the low-redshift universe of $z<0.055$ were used in this study. This is because detecting low surface brightness tidal features in ETGs at higher redshifts is more difficult due to their small angular sizes. Moreover, the detection of tidal features at higher redshifts is also influenced by the effect of the cosmological surface brightness dimming (see Equation (6) in \citealt{YP2022}). We also set a stellar mass cut of $M_\mathrm{star}\ge10^{10}\,M_{\odot}$ because very few ETGs with tidal features were found below this mass limit. When the redshift and stellar mass cuts are applied, the number of ETGs is 212. 

Although the code FIREFLY can recover stellar population properties and star formation histories down to spectra with an S/N of $\sim5$, the recovery of stellar population properties is significantly good for spectra with S/N $\gtrsim20$ \citep{Wilkinson2017}. Therefore, we only used IFU data in which more than 70\% of Voronoi cells have spectra with S/N$\ge10$ in order to ensure the quality of stellar population profiles. Two galaxies are excluded by this criterion. 

Since a large fraction of Voronoi cells in our galaxies have S/N$<20$, requiring more than 70\% of Voronoi cells for a galaxy to have S/N$\ge20$ significantly reduces the number of galaxies in the final sample to 29. We note that the typical uncertainties for the derived age and metallicity of a Voronoi cell are 0.10 dex and 0.13 dex, respectively, for our final sample (see the error bars in the panels (e) and (f) of Figures \ref{fig:ex_1}, \ref{fig:ex_2}, \ref{fig:ex_3}, \ref{fig:ex_4}, and \ref{fig:ex_5}). For the 29 galaxies with high S/Ns, these uncertainties slightly decrease to 0.08 dex and 0.09 dex for age and metallicity, respectively. Therefore, imposing a higher S/N cut on the sample does not substantially reduce the uncertainties of the derived stellar population parameters, but rather only greatly reduces the number of galaxies in the sample.

We also excluded two galaxies that do not have Voronoi cells at $R>1.2R_e$ to ensure that all the ETGs in our sample have a uniform radial coverage up to $R\sim1.5R_e$. So, the number of remaining ETGs is 208. Finally, we excluded an additional 15 galaxies with poor image quality due to their proximity to bright sources (see Section \ref{sec:tf} for a detailed explanation). Therefore, the number of galaxies in the final sample is 193.

The total stellar masses of galaxies used here are from the NASA-Sloan Atlas (NSA) catalog, which is a base catalog for the selection of target galaxies in the MaNGA survey project \citep{Wake2017}. The stellar masses in the NSA catalog were computed from K-correction fits to elliptical Petrosian fluxes. 
\\

\begin{figure*}
\includegraphics[width=\linewidth]{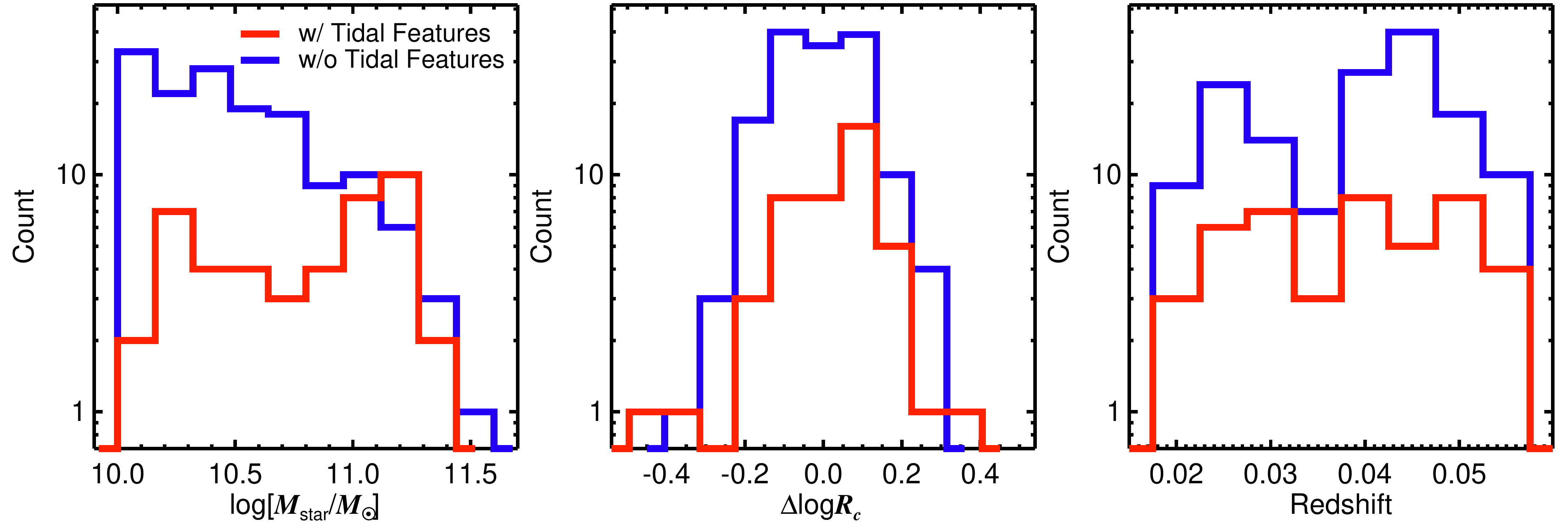}
\caption{Left panel: distributions of $\log(M_\mathrm{star}/M_\odot)$ for our sample. Middle panel: relative distributions of $\log(R_c/\mathrm{kpc})$ ($R_c$: circularized $R_e$) for our sample centered on the mass--size relation of all ETGs ($\Delta\log R_c = \log(R_c/\mathrm{kpc}) - [0.376 + 0.467 (\log(M_\mathrm{star}/M_\odot)-10.5)]$). Right panel: redshift distributions for our sample. In all the panels, ETGs are divided into two categories according to the existence of tidal features (red: with tidal features, and blue: without tidal features).
\label{fig:sam}}
\end{figure*}

\begin{figure*}
\includegraphics[scale=0.30]{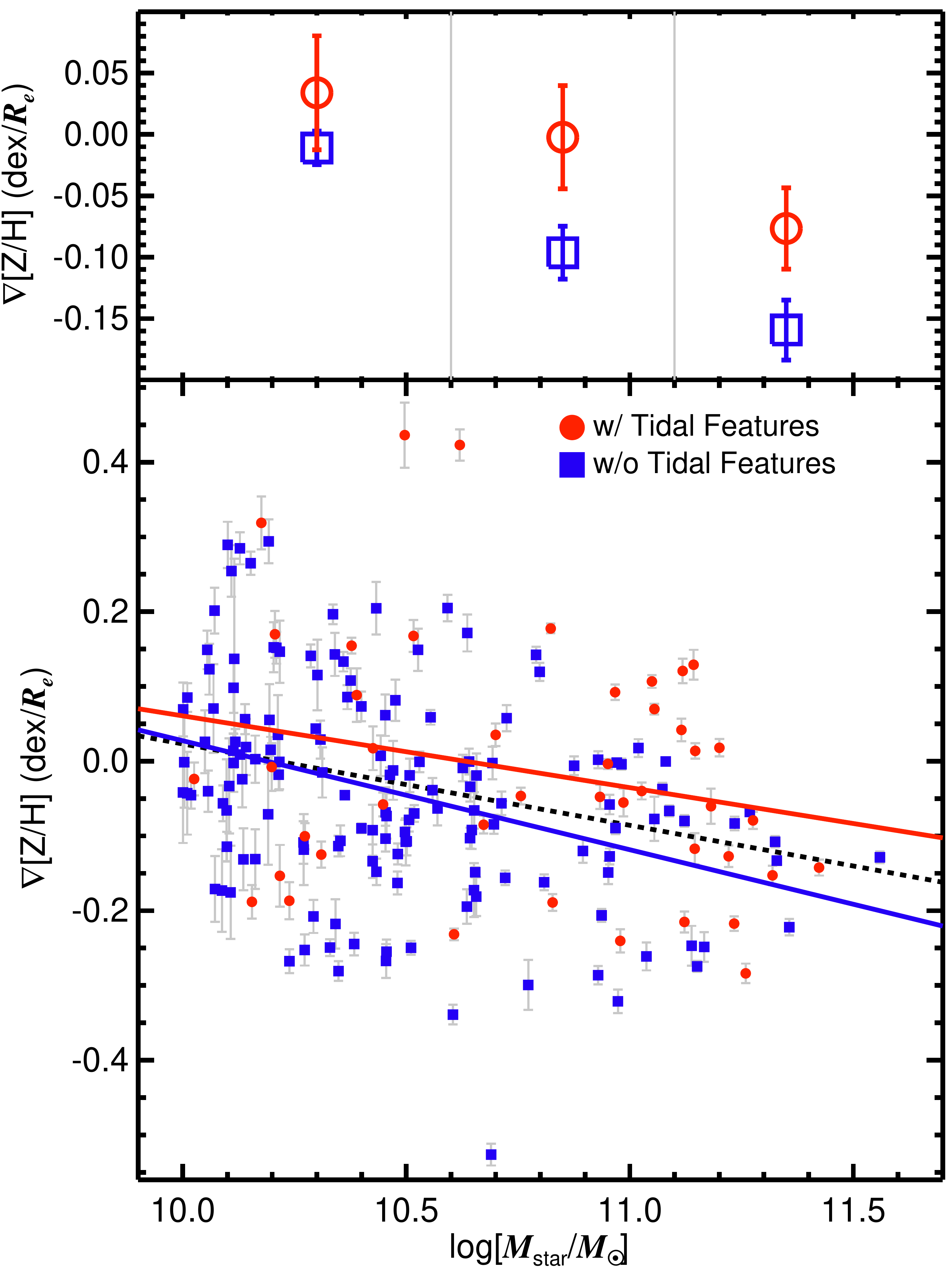}\includegraphics[scale=0.30]{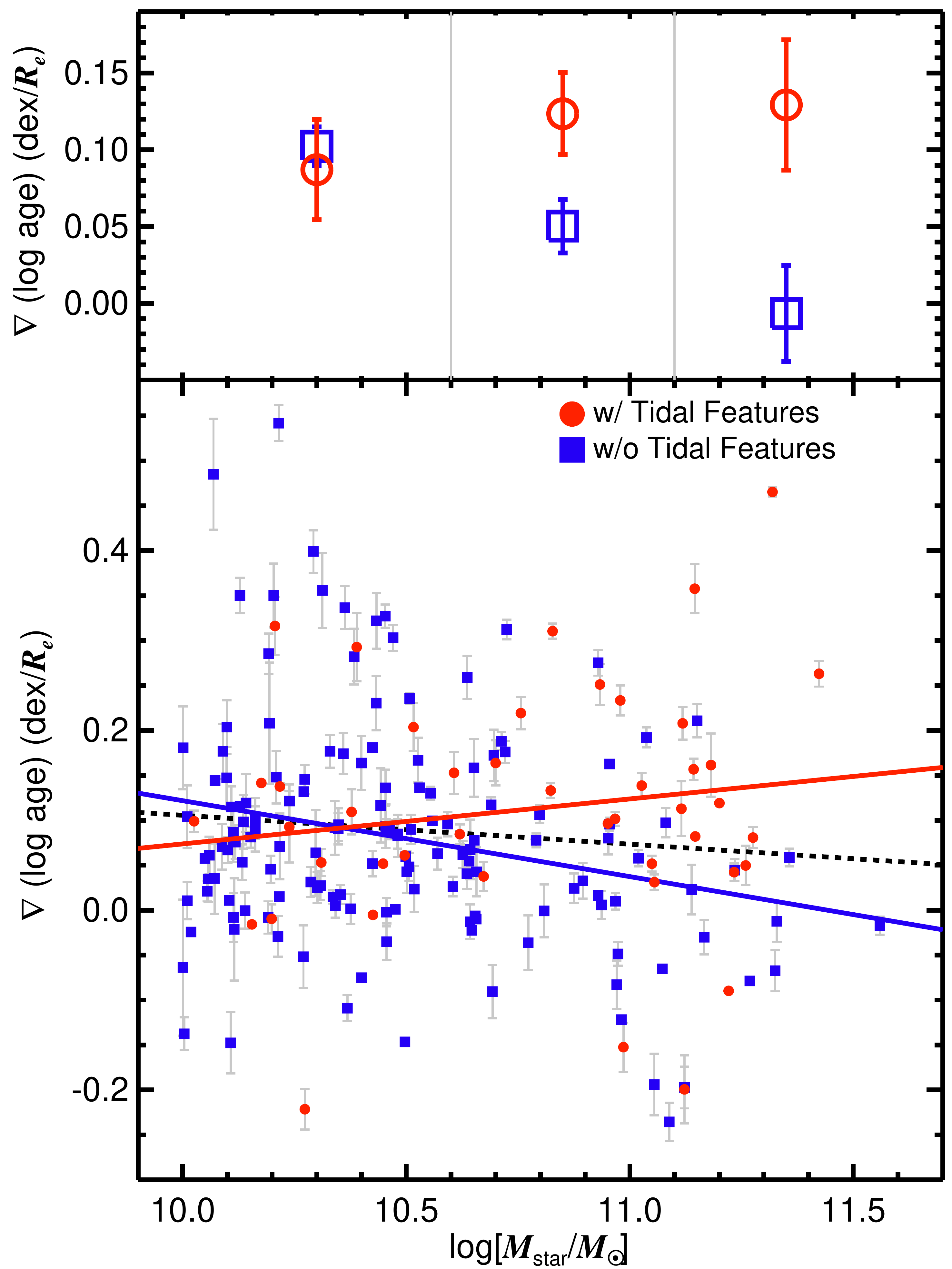}  %width=\linewidth
\centering
\caption{Metallicity gradients (left panel) and age gradients (right panel) of ETGs as a function of galaxy stellar mass. Here, ETGs are divided into two categories according to the existence of tidal features (red circles: with tidal features, and blue squares: without tidal features).  The average values of the gradients within three mass bins and their errors are also displayed at the upper panels. The boundaries of the three mass bins are indicated by the vertical gray lines. The colored solid line is the linear function fitted to ETGs with/without tidal features, while the dashed black line is the linear function for all ETGs (the errors of the data points were taken into account in the linear fittings). See Table \ref{tb:re} for the equations of the linear functions shown in this figure.
\label{fig:gr}}
\end{figure*}

\begin{figure*}
\includegraphics[scale=0.30]{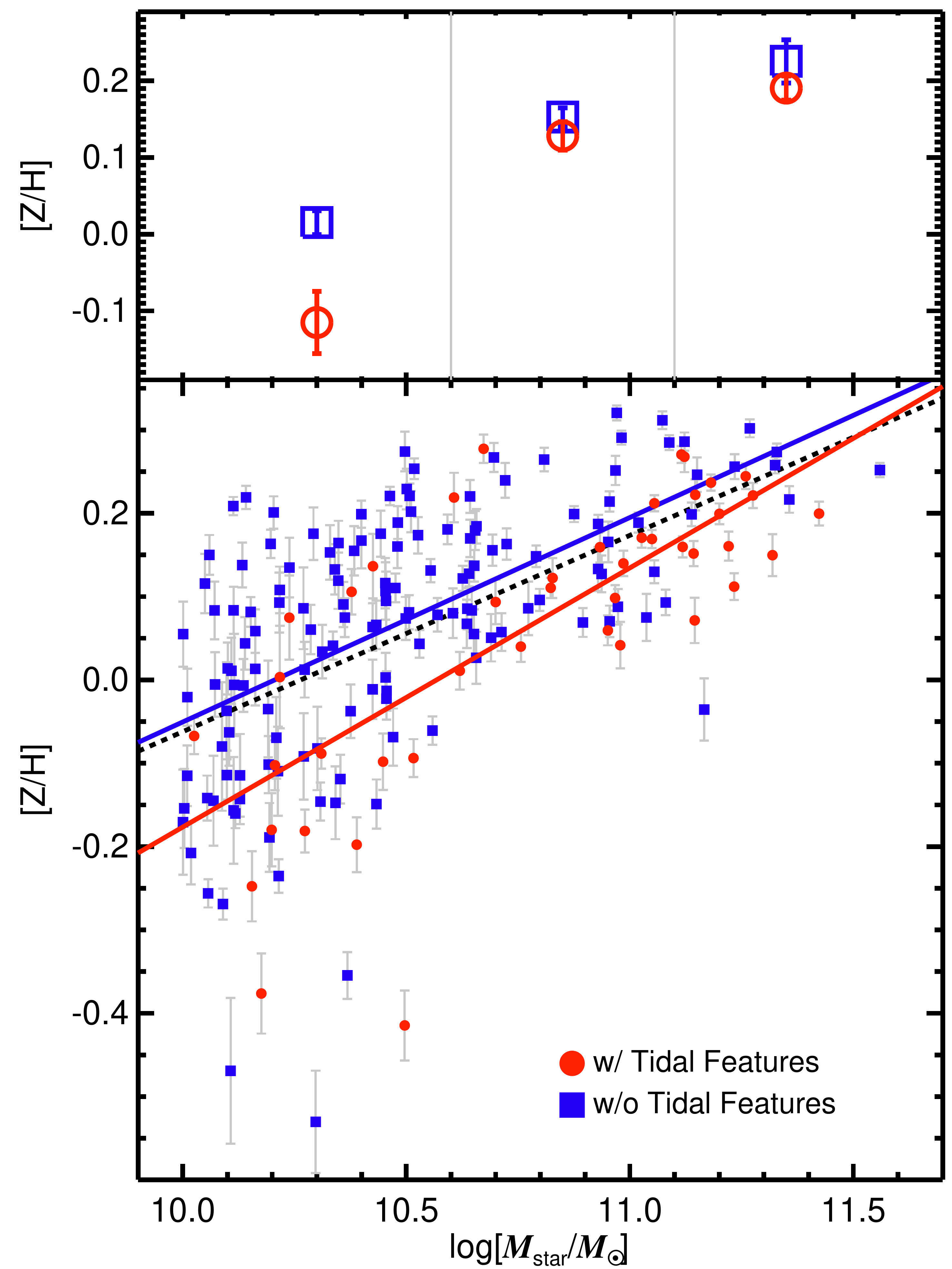}\includegraphics[scale=0.30]{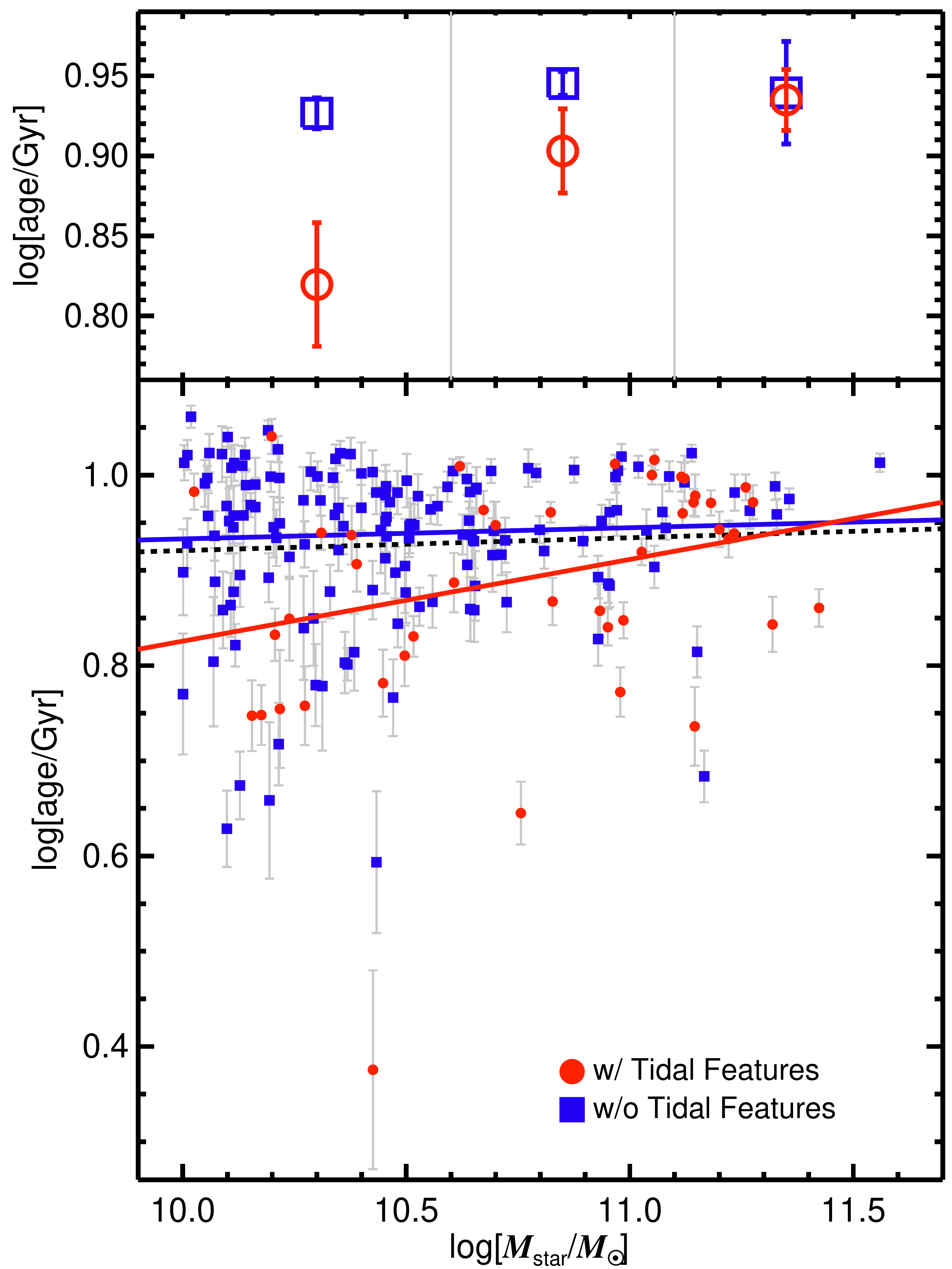}  %width=\linewidth
\centering
\caption{Total metallicities (left panel) and ages (right panel) within $1.5R_e$ of ETGs as a function of stellar mass. ETGs are divided into two categories according to the existence of tidal features (red circles: with tidal features, and blue squares: without tidal features).  The average values of the metallicities and ages within three mass bins and their errors are also displayed at the upper panels. The boundaries of the three mass bins are indicated by the vertical gray lines. The colored solid line is the linear function fitted to ETGs with/without tidal features, while the dashed black line is the linear function for all ETGs (the errors of the data points were taken into account in the linear fittings). See Table \ref{tb:re} for the equations of the linear functions shown in this figure.
\label{fig:tot}}
\end{figure*} 

\subsection{Low Surface Brightness Tidal Features}\label{sec:tf}

To detect tidal features of ETGs, we used coadded images of the Stripe 82 region of the SDSS. The Stripe 82 region, which covers $\sim300$ deg$^2$, was scanned $\sim70$ -- $90$ times during the survey. The repeatedly scanned images can be coadded to make deep images, whose detection limit is $\sim2$ mag deeper than that of single-epoch images of the SDSS \citep{Jiang2014}. Thus, low surface brightness tidal features that are difficult to determine in single-epoch images of the SDSS can be detectable in coadded images of Stripe 82 \citep{Kaviraj2010,Schawinski2010,Hong2015,YL2020,Yoon2022}. 

Here, we used deep coadded images of Stripe 82 from \citet{Jiang2014}. Performing visual inspection on the deep coadded images, we detected tidal features around ETGs as in \citet{YL2020} and \citet{Yoon2022}. We used $r$-band coadded images whose $5\sigma$ detection limit of the aperture magnitude is 24.6, and the surface brightness limit ($1\sigma$ of the background noise over a $1\arcsec\times1\arcsec$ region) is $\sim27$ mag arcsec$^{-2}$. We investigated the deep image of each ETG, adjusting the scale of the pixel values, to detect tidal features such as tails, steams, shells, and diffuse fans. If necessary, we amplified signals by smoothing images using Gaussian kernels with various sizes to identify diffuse faint tidal features more clearly. In the visual inspection process, we excluded 15 ETGs that are too close to bright large galaxies or bright stars because high background levels around the ETGs due to nearby bright sources make it difficult to identify tidal features. The remaining 193 ETGs constitute the final sample of this study.

Examples of deep coadded images of ETGs with tidal features are displayed in panels (b) in Figures \ref{fig:ex_1}, \ref{fig:ex_2}, and \ref{fig:ex_3}. In the figures, tidal features that are not observable in the single-epoch color images are clearly visible in the deep coadded images. Deep images for normal ETGs without tidal features are also shown in panels (b) in Figures \ref{fig:ex_4} and \ref{fig:ex_5}, in which no particular features are detectable around ETGs even in the deep images. We refer to the figures in \citet{YL2020} and \citet{Yoon2022} for more examples of deep images of ETGs with/without tidal features.

We found that 44 of 193 ETGs have tidal features ($22.8\%$). This fraction is much higher than that of \citet{Nair2010}, who used normal SDSS images that are $\sim2$ mag shallower than the deep coadded images used in this study. They found that $\sim3\%$ of ETGs have shell and tail-shaped tidal features, and $\sim6\%$ of ETGs have tidal or disturbed features. Use of images with an extreme depth whose surface brightness limit is deeper than $\sim29$ mag arcsec$^{-2}$ allows us to detect very faint tidal features in most ETGs. For instance, \citet{vanDokkum2005} discovered signatures of tidally disturbed features in $71\%$ of nearby ETGs using such deep images.

In \citet{YL2020} and \citet{Yoon2022}, we have tested our classifications of tidal features by comparing them with the results of \citet{Kaviraj2010}. \citet{Kaviraj2010} classified ETGs with $M_r\footnote{$M_r$: $r$-band absolute magnitude}<-20.5$ at $z<0.05$ in the Stripe 82 region into normal (relaxed) ETGs and ETGs with tidal features. Based on the comparison, we found that more than $\sim90\%$ of their classifications are consistent with ours. We refer to \citet{YL2020} and \citet{Yoon2022} for more detailed information on the test.

We display the distributions of $\log(M_\mathrm{star}/M_\odot)$ for the two samples (ETGs with/without tidal features) in the left panel of Figure \ref{fig:sam}. The distributions show that tidal features are more frequent in more massive ETGs (the average difference in $\log(M_\mathrm{star}/M_\odot)$ between the two samples is $0.29\pm0.07$ dex),\footnote{A Kolmogorov-Smirnov test on the two distributions yields $p=2.1\times10^{-4}$, which means that the two distributions are significantly ($3.7\sigma$) different.} which is consistent with the discoveries of previous studies \citep{Hong2015,YL2020,Yoon2022}.

We compare the galaxy sizes in the two ETG categories in the middle panel of Figure \ref{fig:sam}. Since ETG sizes clearly depend on the stellar mass (mass--size relation of ETGs), the relative distributions of $\log(R_c/\mathrm{kpc})$ ($R_c$: circularized $R_e$) centered on the mass--size relation of all the ETGs in our sample ($\Delta\log R_c = \log(R_c/\mathrm{kpc}) - [0.376 + 0.467 (\log(M_\mathrm{star}/M_\odot)-10.5)]$) are used in the figure. We found that the two distributions are similar to each other: the average values and standard deviations for the $\Delta\log R_c$ distributions of the two samples are almost the same within $1.3$ times the errors.\footnote{A Kolmogorov-Smirnov test on the two distributions gives $p=0.15$, which implies that the two distributions are not significantly different.}

The redshift distributions of the two ETG samples are also displayed in the right panel of Figure \ref{fig:sam}. The two redshift distributions are not significantly different: the average redshift difference between the two samples is 0.0004.\footnote{A Kolmogorov-Smirnov test on the two distributions gives $p=0.44$, which tells us that the two distributions are not significantly different.} The fraction of ETGs with tidal features is nearly unchanged, regardless of the redshift for ETGs with $\log(M_\mathrm{star}/M_\odot)\gtrsim10.7$ ($\sim42\pm6\%$) and ETGs with $\log(M_\mathrm{star}/M_\odot)\lesssim10.7$ at $z\lesssim0.04$ ($\sim20\pm5\%$). The fraction is relatively low in ETGs with $\log(M_\mathrm{star}/M_\odot)\lesssim10.7$ at $z\gtrsim0.04$ ($\sim9\pm3\%$), possibly because depth of the images is not sufficient for detecting tidal features around these ETGs.\footnote{The surface brightness limit of our images, which is $\sim27$ mag arcsec$^{-2}$ in the $r$ band, corresponds to 0.52$L_\odot$ pc $^{-2}$ at $z=0.018$ (the minimum redshift of our ETGs) and 0.60$L_\odot$ pc $^{-2}$ at $z=0.055$. This $15\%$ difference is owing to the cosmological dimming effect.} However, even after dividing our ETG sample into two redshift ranges ($z<0.04$ and $z\ge0.04$), the results in Section \ref{sec:results} are still found in both the redshift ranges. Therefore, even if the possible redshift bias really exists, it is not serious and cannot essentially affect our conclusion.
\\

\section{Results}\label{sec:results}
We display $\nabla[Z/H]$ and $\nabla(\log\mathrm{age})$ of ETGs as a function of stellar mass in Figure \ref{fig:gr} (see Table \ref{tb:re} for the equations of the linear functions shown in the figure). As shown in the left panel of the figure, ETGs have a negative $\nabla[Z/H]$ of $-0.04$ dex$/R_e$ (lower $[Z/H]$ in outer parts) on average. The figure also shows that more massive ETGs have a steeper negative $\nabla[Z/H]$ (the slope of the mass dependence is $-0.109\pm0.027$ for all ETGs). So, the mean $\nabla[Z/H]$ is -0.11 dex$/R_e$ for ETGs with $\log(M_\mathrm{star}/M_\odot)\ge11.1$, while that of ETGs with $10.0\le\log(M_\mathrm{star}/M_\odot)<10.6$ is -0.01 dex$/R_e$. Interestingly, ETGs with tidal features have shallower (or higher) $\nabla[Z/H]$ on average than ETGs without tidal features by $\sim0.04$ -- $0.09$ dex$/R_e$ for a given stellar mass. 

We conducted 1,000,000 random resamplings for our ETGs with $\log(M_\mathrm{star}/M_\odot)>10.6$ and found that ETGs with tidal features in $\sim996,500$ resampled sets have higher mean values of $\nabla[Z/H]$ than ETGs without tidal features for a given mass. This means that the statistical significance is $99.65\%$ ($\sim2.9\sigma$) for the fact that ETGs with tidal features at $\log(M_\mathrm{star}/M_\odot)\ge10.6$ have higher $\nabla[Z/H]$ on average than the counterparts without tidal features for a given stellar mass. We also tested the significance of the difference between the two categories (ETGs with/without tidal features) by conducting a Kolmogorov-Smirnov (KS) test on the relative distributions (deviations) of $\nabla[Z/H]$ centered on the linear function of all ETGs (the black dashed line in the left panel of Figure \ref{fig:gr}). It is found that the probability ($0\le p\le1$) of the null hypothesis in which the two distributions are drawn from the same distribution is $p=1.67\times10^{-2}$ at $\log(M_\mathrm{star}/M_\odot)\ge10.45$ ($\sim2.4\sigma$ significance in the difference).

The right panel of Figure \ref{fig:gr} shows that ETGs have a positive $\nabla(\log\mathrm{age})$ of $0.09$ dex$/R_e$ on average. The mass dependence in $\nabla(\log\mathrm{age})$ for all ETGs is not significant (the slope of the relation is $-0.032\pm0.024$). In this case, ETGs with tidal features have more positive $\nabla(\log\mathrm{age})$ on average than ETGs without tidal features, except for ETGs at $\log(M_\mathrm{star}/M_\odot)\lesssim10.6$, in which both the ETG populations have similar mean values of $\nabla(\log\mathrm{age})$ ($\sim0.1$ dex$/R_e$). The difference in $\nabla(\log\mathrm{age})$ is larger for more massive galaxies, so that at $\log(M_\mathrm{star}/M_\odot)\ge11.1$, the mean $\nabla(\log\mathrm{age})$ is 0.13 dex$/R_e$ for ETGs with tidal features, whereas that of ETGs without tidal features is -0.01 dex$/R_e$. 

Conducting the random resamplings, we found that the statistical significance is $99.88\%$ ($\sim3.2\sigma$) for the fact that ETGs with tidal features at $\log(M_\mathrm{star}/M_\odot)\ge10.6$ have higher $\nabla(\log\mathrm{age})$ on average than the counterparts without tidal features for a given mass. From the KS test on the distributions (deviations) of $\nabla(\log\mathrm{age})$ relative to the linear function of all ETGs, we found $p=1.27\times10^{-3}$ for the two ETG categories at $\log(M_\mathrm{star}/M_\odot)\ge10.45$ (i.e., $\sim3.2\sigma$ significance in the difference).

The general consensus of previous studies on the stellar population gradients of ETGs is that ETGs have negative metallicity gradients and positive/flat age gradients \citep{Annibali2007,Koleva2011,Cook2016,Goddard2017,SanRoman2018,Lacerna2020,Neumann2021}, which is consistent with our results. Steeper negative metallicity gradients in more massive ETGs are also widely found in several recent works \citep{Forbes2005,Tortora2011,Goddard2017,MartinNavarro2018,Lacerna2020,Neumann2021}.

In Figure \ref{fig:tot} we display the total metallicities and ages of ETGs calculated within $1.5R_e$ as a function of stellar mass (see Table \ref{tb:re} for the equations of the linear functions shown in the figure). The left panel of Figure \ref{fig:tot} illustrates the mass--metallicity relation of ETGs, which shows that higher-mass ETGs have higher metallicities (the slope of the relation is $0.24\pm0.02$). The figure also shows that ETGs with tidal features have lower $[Z/H]$ on average than ETGs without tidal features by $\sim0.07$ dex for a given mass. The difference in $[Z/H]$ between the two populations is $0.03$ dex at $\log(M_\mathrm{star}/M_\odot)\ge10.6$, and it is $0.13$ dex at $\log(M_\mathrm{star}/M_\odot)<10.6$. 

The statistical significance for the fact that ETGs with tidal features have lower $[Z/H]$ on average than ETGs without tidal features for a given mass is $99.98\%$ ($\sim3.7\sigma$), but it reduces to $98.5\%$ for ETGs at $\log(M_\mathrm{star}/M_\odot)\ge10.6$. For the two ETG categories, we found $p=1.26\times10^{-4}$ from the KS test on the relative distributions (deviations) of $[Z/H]$ from the linear function of all ETGs (i.e., $\sim3.8\sigma$ significance in the difference).

The right panel of Figure \ref{fig:tot} shows that the mass--age relation of all ETGs or ETGs without tidal features is almost flat (the slope is $0.01\pm0.02$), which means that most of them are dominated by old stellar populations whose ages exceed $\sim7$ Gyr, regardless of stellar mass. The figure shows that ETGs with tidal features at $\log(M_\mathrm{star}/M_\odot)\lesssim11.1$ have younger stellar population ages on average than ETGs without tidal features by $\sim1$ -- $2$ Gyr, whereas the age difference between the two categories is negligible in the most massive ETGs with $\log(M_\mathrm{star}/M_\odot)\gtrsim11.1$. 

The statistical significance is $99.94\%$ ($\sim3.4\sigma$) for the fact that ETGs with tidal features at $\log(M_\mathrm{star}/M_\odot)<11.1$ have younger stellar populations on average than the counterparts without tidal features for a given stellar mass. The KS test on the relative distributions (deviations) of $\log\mathrm{age}$ centered on the linear function of all ETGs gives $p=2.46\times10^{-4}$ for the two ETG categories at $\log(M_\mathrm{star}/M_\odot)<11.0$ (i.e., $\sim3.7\sigma$ significance in the difference).

\begin{deluxetable*}{rlll}
\tablecaption{Equations for Relations between Stellar Population Properties and Stellar Mass\label{tb:re}}
\tablehead{\colhead{} & \colhead{ETGs with Tidal Features} & \colhead{ETGs without Tidal Features} & \colhead{All ETGs}
}
\startdata
$\nabla[Z/H]=\alpha+\beta[\log(M_\mathrm{star}/M_\odot)-10.5]$&$\alpha=0.013\pm0.030$&$\alpha=-0.046\pm0.011$&$\alpha=-0.031\pm0.010$\\
&$\beta=-0.096\pm0.062$&$\beta=-0.146\pm0.031$&$\beta=-0.109\pm0.027$\\
\hline
$\nabla(\log\mathrm{age})=\alpha+\beta[\log(M_\mathrm{star}/M_\odot)-10.5]$&$\alpha=0.099\pm0.025$&$\alpha=0.080\pm0.010$&$\alpha=0.089\pm0.009$\\
&$\beta=0.050\pm0.052$&$\beta=-0.084\pm0.028$&$\beta=-0.032\pm0.024$\\
\hline
$[Z/H]=\alpha+\beta[\log(M_\mathrm{star}/M_\odot)-10.5]$&$\alpha=-0.021\pm0.022$&$\alpha=0.072\pm0.010$&$\alpha=0.056\pm0.009$\\
&$\beta=0.311\pm0.044$&$\beta=0.245\pm0.028$&$\beta=0.236\pm0.023$\\
\hline
$\log(\mathrm{age/Gyr})=\alpha+\beta[\log(M_\mathrm{star}/M_\odot)-10.5]$&$\alpha=0.869\pm0.020$&$\alpha=0.939\pm0.006$&$\alpha=0.928\pm0.006$\\
&$\beta=0.086\pm0.041$&$\beta=0.012\pm0.018$&$\beta=0.013\pm0.016$\\
\enddata
\tablecomments{The unit of stellar population gradients is dex/$R_e$.
}
\end{deluxetable*}

\begin{figure*}
\includegraphics[width=\linewidth]{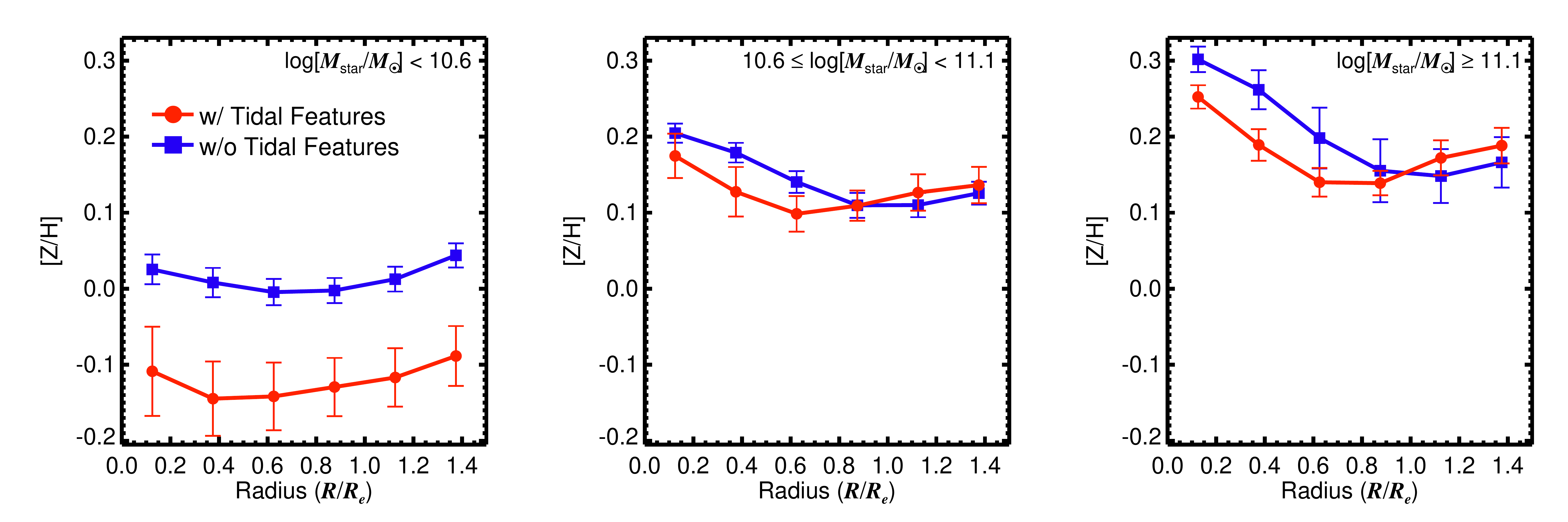}
\includegraphics[width=\linewidth]{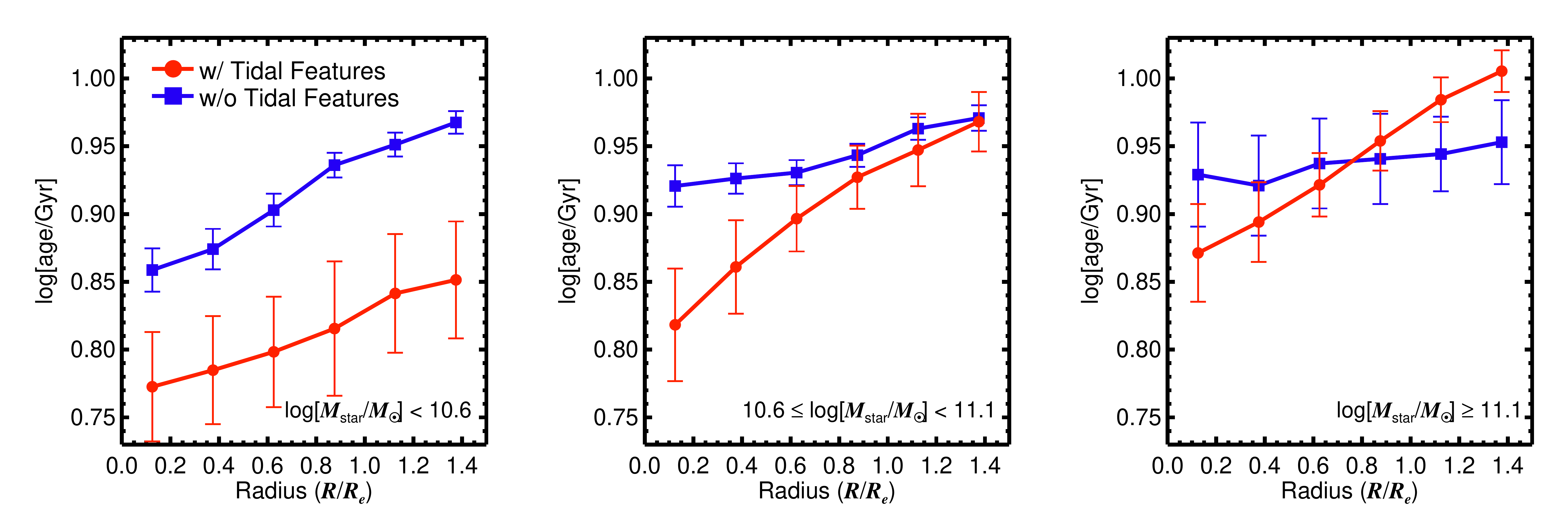}  %width=\linewidth
\caption{Upper panels: mean metallicity profiles for ETGs in different mass bins. Lower panels: mean age profiles for ETGs in different mass bins. The error bar indicates the standard deviation of the mean values in each bin computed from 1000 bootstrap resamplings. ETGs are divided into two categories according to the existence of tidal features (red circles: with tidal features, and blue squares: without tidal features). The number of ETGs with tidal features included in each of the mass bins is 15 at $\log(M_\mathrm{star}/M_\odot)<10.6$, 15 at  $10.6\le\log(M_\mathrm{star}/M_\odot)<11.1$, and 14 at $\log(M_\mathrm{star}/M_\odot)\ge11.1$. The number of ETGs without tidal features in each bin is 98 at $\log(M_\mathrm{star}/M_\odot)<10.6$, 41 at  $10.6\le\log(M_\mathrm{star}/M_\odot)<11.1$, and 10 at $\log(M_\mathrm{star}/M_\odot)\ge11.1$.
\label{fig:profile}}
\end{figure*} 

\begin{figure*}
\includegraphics[width=\linewidth]{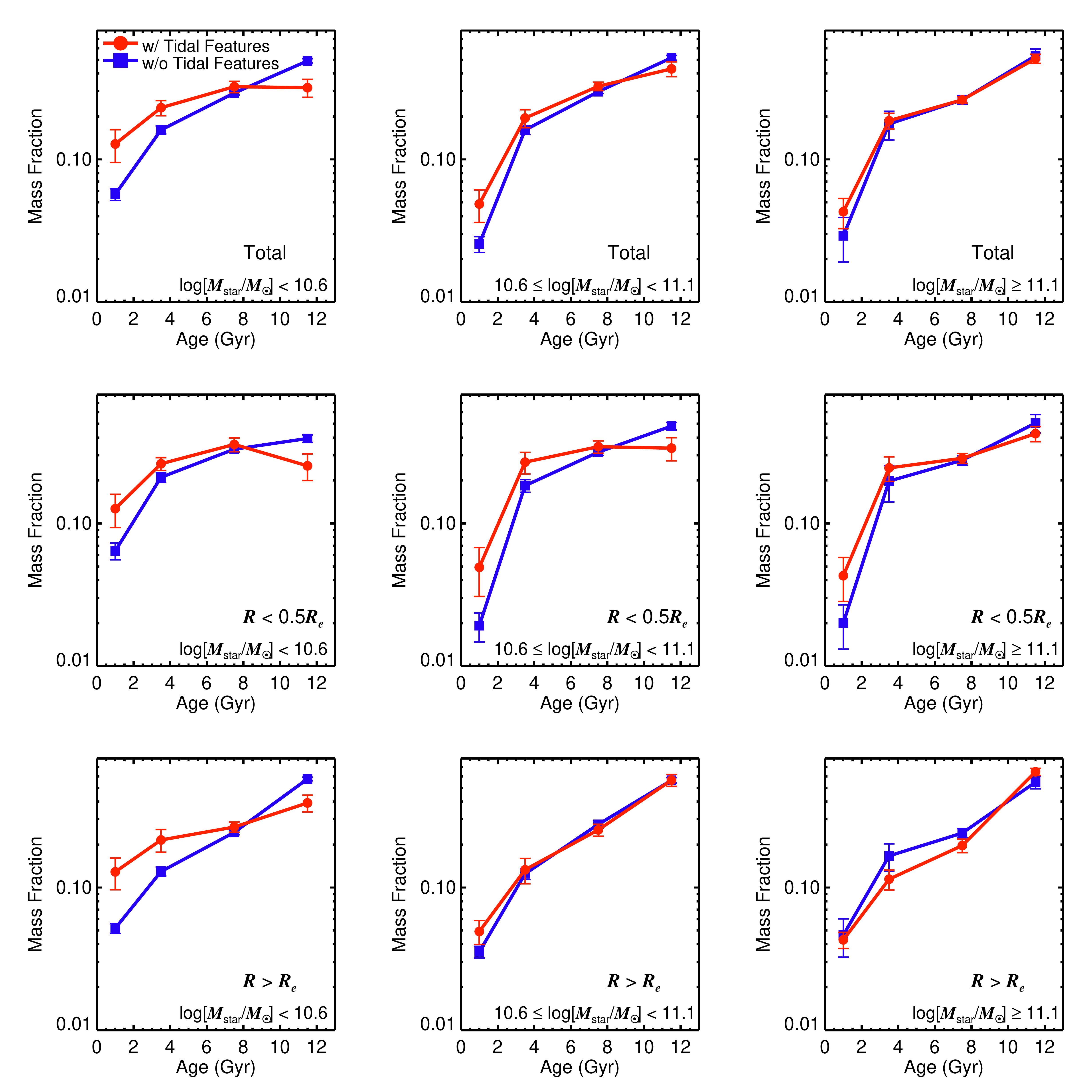}
\centering
\caption{Average mass fraction as a function of stellar population age, which represents the average star formation history of ETGs. Here, the mass fraction means the mass of the whole SSPs in each age bin divided by the total mass of all the SSPs of a galaxy. The four age bins are defined as age/Gyr $<2$, $2\le$ age/Gyr $<5$, $2\le$ age/Gyr $<5$, $5\le$ age/Gyr $<10$, and age/Gyr $\ge10$. Each error bar denotes the standard deviation of the mean values from 1000 bootstrap resamplings. ETGs are divided into two categories according to the existence of tidal features (red circles: with tidal features, and blue squares: without tidal features). In this figure, we display the results for ETGs in different mass bins (the panels on the right show the results for more massive ETGs). The top panels show the results for stellar populations in the total areas ($R<1.5R_e$). The panels in the second row are for the central regions of $R<0.5R_e$, and the bottom panels are for the outer regions of $R>R_e$.
\label{fig:sfh}}
\end{figure*} 

\begin{figure*}
\includegraphics[width=\linewidth]{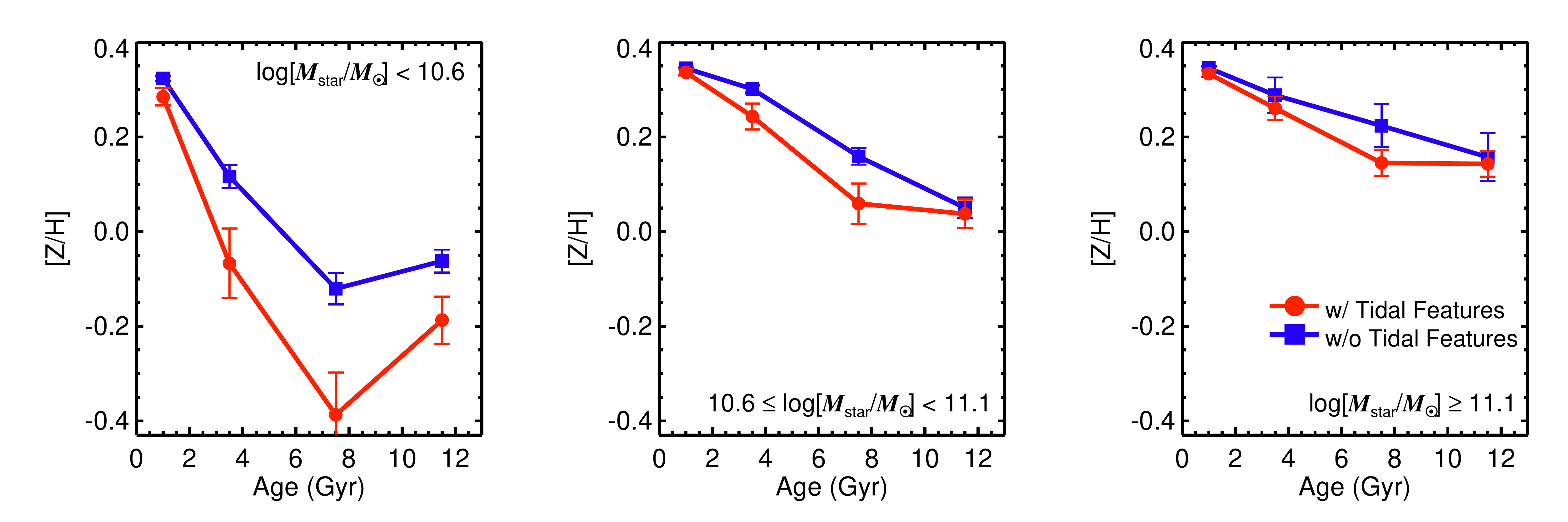}
\centering
\caption{Average metallicity as a function of stellar population age, which corresponds to the average metallicity enrichment history of the ETGs. The four age bins are defined as age/Gyr $<2$, $2\le$ age/Gyr $<5$, $2\le$ age/Gyr $<5$, $5\le$ age/Gyr $<10$, and age/Gyr $\ge10$. Each error bar indicates the standard deviation of the mean values from 1000 bootstrap resamplings. The ETGs are divided into two categories according to the existence of tidal features (red circles: with tidal features, and blue squares: without tidal features). Here, we show the results for ETGs in different mass bins (the panel on the right shows the result for more massive ETGs). 
\label{fig:met}}
\end{figure*} 

Figure \ref{fig:profile} displays the mean metallicity/age profiles for ETGs in different mass bins. The profiles in the figure were derived by dividing the spaxels in the IFU data of ETGs into six radial bins (normalized by $R_e$) and calculating the average metallicity/age values within the radial bins. All the major trends of Figures \ref{fig:gr} and \ref{fig:tot} described below are also clearly seen in Figure \ref{fig:profile}. 
\begin{itemize}
\item{ETGs with $\log(M_\mathrm{star}/M_\odot)\gtrsim10.6$ have negative $\nabla[Z/H]$.}
\item{ETGs with tidal features have shallower $\nabla[Z/H]$ than ETGs without tidal features at $\log(M_\mathrm{star}/M_\odot)\gtrsim10.6$.}
\item{ETGs with tidal features have lower $[Z/H]$ than ETGs without tidal features.}
\item{ETGs with tidal features have more positive $\nabla(\log\mathrm{age})$ than ETGs without tidal features at $\log(M_\mathrm{star}/M_\odot)\gtrsim10.6$.}
\item{ETGs with tidal features at $\log(M_\mathrm{star}/M_\odot)\lesssim11.1$ have younger stellar populations than the counterparts without tidal features.}
\end{itemize}

We examined the star formation histories of our ETGs using the information of mass weights of individual SSPs contributing to each spaxel. All SSPs contributing to each spaxel of a galaxy were divided into four stellar population age bins: age/Gyr $<2$, $2\le$ age/Gyr $<5$, $2\le$ age/Gyr $<5$, $5\le$ age/Gyr $<10$, and age/Gyr $\ge10$. Then, we derived the mass fraction of each age bin, which is the mass of the whole SSPs in each age bin divided by total mass of all the SSPs of a galaxy. The mass fraction as a function of the age bin represents the star formation history of a galaxy. By calculating the average mass fraction of each age bin for multiple ETGs, we obtained the average star formation history of ETGs. 

Figure \ref{fig:sfh} displays the average mass fractions as a function of stellar population age (the average star formation histories) for ETGs with/without tidal features at different stellar mass bins. In the figure, we also show the results for the stellar populations in central regions ($R<0.5R_e$) and outer regions ($R>R_e$). Figure \ref{fig:sfh} shows that $\sim80\%$ of the stellar populations in ETGs are older than 5 Gyr. Interestingly, ETGs with tidal features have a higher fraction of stars that were recently formed within 5 Gyr than ETGs without tidal features: $5\pm2$ and $14\pm4$ percentage points higher at $\log(M_\mathrm{star}/M_\odot)\geq10.6$ and $\log(M_\mathrm{star}/M_\odot)<10.6$, respectively. The stellar population fractions at age $<2$ Gyr are also higher in ETGs with tidal features: $2\pm1$ and $7\pm3$ percentage points higher at $\log(M_\mathrm{star}/M_\odot)\geq10.6$ and $\log(M_\mathrm{star}/M_\odot)<10.6$, respectively. 

In the case of ETGs with $\log(M_\mathrm{star}/M_\odot)\geq10.6$, the trend that ETGs with tidal features have a higher fraction of recently formed stars is clearly found in the central regions of $R<0.5R_e$ in the sense that the fraction of stellar populations with an age $<5$ Gyr is $10\pm4$ percentage points higher in ETGs with tidal features than in ETGs without tidal features. By contrast, the trend is insignificant or absent in the outer regions of $R>R_e$.

 In the case of lower-mass ETGs at $\log(M_\mathrm{star}/M_\odot)<10.6$, the trend that ETGs with tidal features have a higher fraction of young stellar populations is not only found in the central regions, but also in the outer regions of galaxies: in both regions, the fraction of stellar populations with an age $<5$ Gyr is more than $11\pm5$ percentage points higher in ETGs with tidal features than in ETGs without tidal features. 

We also investigated the metallicity enrichment histories for the stellar populations of our ETGs. As in the process of deriving the star formation histories of the ETGs described above, we separated all SSPs for each spaxel of a galaxy into four age bins. The metallicity $[Z/H]$ as a function of the age bin, which corresponds to the metallicity enrichment history of a galaxy, was derived by calculating mass-weighted $[Z/H]$ within each age bin. By computing the average $[Z/H]$ of each age bin for multiple ETGs, an average metallicity enrichment history of ETGs was obtained.

Figure \ref{fig:met} shows the average metallicities as a function of stellar population age (the average metallicity enrichment history) for ETGs with/without tidal features in different stellar mass bins. The figure suggests that the oldest stellar populations with an age $>10$ Gyr in more massive galaxies have higher metallicities ($[Z/H]\sim0.15$ for ETGs with $\log(M_\mathrm{star}/M_\odot)\ge11.1$ and $[Z/H]\sim-0.1$ for ETGs with $\log(M_\mathrm{star}/M_\odot)<10.6$), which is consistent with the fact that the mass--metallicity relation is already present in the early universe at $z\gtrsim2$ \citep{Maiolino2008,Mannucci2009,Zahid2013,Zahid2014}. The general metal-enrichment history of our ETGs is that more recently formed stars have higher metallicities than stars that formed in the early universe. 

Figure \ref{fig:met} shows that stellar populations with an age $\sim7$ Gyr in ETGs with tidal features have lower metallicities by $\sim0.1$ -- $0.3$ dex than those in ETGs without tidal features, so that ETGs with tidal features have slower (tardy) metal-enrichment histories than ETGs without tidal features in the high-redshift universe. However, the metallicities of young stellar populations in ETGs with tidal features, especially the metallicities of young stars with an age $<2$ Gyr, are close to those of ETGs that do not have tidal features (e.g., $[Z/H]\sim0.34$ for ETGs with $\log(M_\mathrm{star}/M_\odot)>10.6$). Combined with the fact that ETGs with tidal features have a higher fraction of young stellar populations, it follows that ETGs with tidal features have been speeding up the metal enrichment through stars that formed or were accreted in recent merger processes.
\\

\section{Discussion}\label{sec:discuss}

In Figure \ref{fig:profile}, ETGs, especially with $\log(M_\mathrm{star}/M_\odot)\ge10.6$, have a nonlinear metallicity profile on average: the inner profile at $R\lesssim0.7R_e$ is steeply negative, and the outer profile at  $R\gtrsim0.7R_e$ is flat (or slightly positive).\footnote{Specifically, negative metallicity profiles become flat (or even positive) at $R>0.6R_e$ for ETGs with tidal features, whereas they flatten at $R>0.9$ -- $1.1R_e$ for ETGs without tidal features.} The nonlinear metallicity profiles of ETGs were also detected in previous studies \citep{Oyarzun2019,Lacerna2020,Zibetti2020}. The nonlinear shape of the metallicity profiles is evidence of the two-phase formation scenario, which suggests that ETGs rapidly form through gas-rich collapse/mergers with massive star formations in the early universe beyond $z\sim2$ and grow gradually by multiple dry mergers in the lower-redshift universe \citep{Naab2009,Oser2010,Taylor2017,Oyarzun2019,Lacerna2020,Zibetti2020}.

Gas-rich mergers, which are the first phase of the two-phase formation scenario, can induce gas inflows into the central parts of galaxies due to loss of angular momentum through gravitational torques and radiation, and then the accumulated gas in the centers of galaxies forms young stars \citep{Hernquist1989,Barnes1991,Barnes1996,Hopkins2008a}. If the amount of (cold) gas in galaxies is very high, as in the early universe, efficient star formation intensively occurs there, leading to rapid metal enrichment \citep{Kennicutt1998,Kobayashi2004,Hopkins2009,Cook2016}. Since the inner regions of galaxies have high/steep gravitational potentials due to the concentrated baryonic mass, galactic winds from supernovae are not easily able to blow away the enriched metallicities \citep{Cook2016}. 

At the same time, the star formation efficiency is relatively low in the outer regions of galaxies due to the low gas density. Moreover, the outer parts of galaxies are environments where gravitational potentials are not deep, so that galactic winds from supernovae can scatter the metallicities more easily than in the central regions \citep{Cook2016}. So, the metallicity enrichment in the outer regions of galaxies cannot be as efficient as in the central parts. Therefore, gas-rich collapse/mergers in the early universe (the first phase in the two-phase formation scenario) can create steeply negative metallicity profiles \citep{Kobayashi2004,Forbes2005,Spolaor2009,Tortora2011,Cook2016,Taylor2017,Oyarzun2019}, such as those found in the inner parts of the metallicity profiles ($R\lesssim0.7R_e$) of our ETGs (Figure \ref{fig:profile}).\footnote{The second formation phase at lower redshifts, which is galaxy accretions, cannot affect the inner regions of galaxies easily, so that the steeply negative metallicity gradients in the central parts can be maintained even in the low-redshift universe.}

More massive galaxies are less affected by galactic winds\footnote{In low-mass galaxies, the metallicity gradients can be flat or even positive through galactic winds \citep{Mori1997,Taylor2017}.} and can more efficiently retain large amounts of metal-enriched gas in the central parts due to their deep gravitational potentials\footnote{Stars are more concentrated in higher-mass ETGs \citep{Ferrarese2006,Kormendy2009,Savorgnan2013}.} \citep{Mori1997,Tortora2011,Taylor2017}, which naturally explains our finding that metallicity gradient is more steeply negative in higher-mass galaxies, as shown in Figure \ref{fig:gr}.

The second formation phase is multiple mergers in the lower-redshift universe. The amount of (cold) gas in galaxies is generally lower at lower redshifts \citep{Erb2006,Daddi2010,Tacconi2010,Geach2011,Carilli2013,Morokuma2015,Yoon2019} because galaxies gradually consume the available gas in diverse ways throughout the history of the universe. So, mergers that occurred in the lower-redshift universe, particularly in the local universe, are expected to be more dissipationless (dry) than those in the early universe. These mergers/accretions deposit stars mainly in the outer parts of galaxies and can flatten the outer metallicity profiles by redistributing/mixing stars \citep{Kobayashi2004,Ko2005,DiMatteo2009,Lackner2012,Hirschmann2015,Cook2016,Taylor2017,Ferreras2019,Oyarzun2019}, as seen in our results for the metallicity profiles of ETGs with $\log(M_\mathrm{star}/M_\odot)\ge10.6$ shown in Figure \ref{fig:profile}. This can also explain our finding that ETGs with tidal features at $\log(M_\mathrm{star}/M_\odot)\gtrsim10.6$, which are likely to have experienced recent mergers, have shallower metallicity gradients than the counterparts without tidal features. 

If we examined metallicities of stellar populations in very outer regions (e.g., $R\gtrsim3$ -- $5R_e$) of ETGs, which are beyond the spatial coverage used here, and included them when we derive metallicity gradients, our result might be different in the sense that ETGs have shallower metallicity gradients than those shown in Section \ref{sec:results}, particularly for massive ETGs. This is because the inner regions of ETGs, where the imprints of the first gas-rich formation phase are still dominant, are reduced proportionally as the coverage is extended, and massive galaxies are expected to be formed through multiple dry mergers \citep{Desroches2007,Bernardi2011a,Bernardi2011b,Yoon2017,YL2020,OLeary2021} that can flatten the metallicity profiles in the outer part of galaxies.

In the processes of mergers that occurred in the low-redshift universe, small amounts of gas can fall into the central parts of galaxies, which results in the formation of young stars \citep{Hernquist1989,Barnes1991,Barnes1996,Hopkins2008a}. Meanwhile, ages of accreted/deposited stars in the outer regions by mergers may be relatively old \citep{Lackner2012,Hirschmann2015}. This can explain the steeply positive age profiles in ETGs with tidal features (Figures \ref{fig:gr} and \ref{fig:profile}). As shown in \citet{Hopkins2009}, the steeply positive age profile caused by a recent merger flattens out within a few gigayears if the stars in the galaxies age uniformly (passive evolution).

Low-mass ETGs with $\log(M_\mathrm{star}/M_\odot)\lesssim10.6$ are more susceptible to galaxy mergers due to their low masses and have fewer capabilities to hold gas in the centers than higher-mass ETGs. Thus, recent mergers are able to impact entire regions of these galaxies in terms of star formation activity. This may explain the fact that ETGs with tidal features at $\log(M_\mathrm{star}/M_\odot)\lesssim10.6$ have younger stellar populations throughout the galaxies than the counterparts without tidal features (Figures \ref{fig:tot} and \ref{fig:profile}).

Unlike ETGs with tidal features that are very likely to have undergone recent mergers a few gigayears ago \citep{Ji2014,Mancillas2019,YL2020}, ETGs without tidal features may not have experienced significant mergers within the last few gigayears. Instead, ETGs without tidal features probably grew and reached their current masses and metallicities in the earlier universe, in which gas-rich processes/mergers that can efficiently enrich the metallicity are more dominant. Recent mergers that ETGs with tidal features experienced can also increase the metallicities, as mentioned in Section \ref{sec:results} with Figure \ref{fig:met}. However, in the low-redshift universe, the gas abundance is relatively low, so that metal enrichment through the formation of new stars in merger processes is not effective compared with that in the early universe. Unlike the metallicity, the stellar mass can be increased relatively easily through the recent merger process that combines the stellar masses of the progenitor galaxies. Therefore, it is natural that the metallicities of ETGs with tidal features are similar to those of less massive normal ETGs. According to the mass--metallicity relation, this means that ETGs with tidal features have lower metallicities than normal ETGs, as we found in this study (Figures \ref{fig:tot} and \ref{fig:profile}). This explanation applies more prominently to low-mass ETGs at $\log(M_\mathrm{star}/M_\odot)\lesssim10.6$ (hence, the metallicity gap between ETGs with and without tidal features is large) because the same merger can boost the stellar masses by a greater proportion for lower-mass galaxies. 
\\

\section{Summary}\label{sec:summary}

We investigated the impact of galaxy mergers on the stellar population (age and metallicity) profiles/gradients of ETGs. Our sample is constructed from ETGs in the Stripe 82 region of the SDSS with MaNGA IFU spectroscopy data. The ETGs in our sample lie at $z<0.055$ and their number is 193. In this study, spatially resolved stellar population properties are from the MaNGA FIREFLY VAC based on the final release version of the SDSS. Tidal features around ETGs, which are direct evidence for recent mergers, were detected through visual inspection of deep coadded images of the Stripe 82 region that are $\sim2$ mag deeper than single-epoch images of the SDSS. Our main results are as follows. Note that all the descriptions below indicate average properties.

\begin{enumerate}
\item ETGs with $\log(M_\mathrm{star}/M_\odot)\gtrsim10.6$ have negative metallicity gradients (lower metallicity in the outer parts; Figures \ref{fig:gr} and \ref{fig:profile}).

\item ETGs with tidal features have shallower metallicity gradients by $\sim0.08$ dex$/R_e$ and more positive age gradients by $\sim0.09$ dex$/R_e$ than ETGs without tidal features at $\log(M_\mathrm{star}/M_\odot)\gtrsim10.6$ (Figures \ref{fig:gr} and \ref{fig:profile}). 

\item When all the resolved stellar populations are integrated, ETGs with tidal features have lower metallicities by $\sim0.07$ dex and younger ages by $\sim1$ -- $2$ Gyr than ETGs without tidal features (Figure \ref{fig:tot}). 

\item In the central regions ($R<0.5R_e$) of galaxies, the mass fraction of young stellar populations with an age $< 5$ Gyr is $\sim10$ percentage points higher in ETGs with tidal features than in ETGs without tidal features (Figure \ref{fig:sfh}).

\item ETGs with tidal features have slower (tardy) metal-enrichment histories than ETGs without tidal features in the early universe. However, ETGs with tidal features have been increasing their stellar metallicities more rapidly than ETGs without tidal features over the last $\sim5$ Gyr (Figure \ref{fig:met}).
\end{enumerate}
We note that ETGs with tidal features can be directly translated into ETGs that have recently experienced galaxy mergers in the main results described above. 

We discussed several influences of galaxy mergers on stellar population profiles based on our results and related studies. 
\begin{enumerate}
\item The negative metallicity profiles of ETGs in the central regions can be traces of gas-rich mergers that occurred in the early universe at $z\gtrsim2$.

\item Recent mergers, which are expected to be more dissipationless than those in the early universe, can flatten the metallicity profiles and are likely to have lower capabilities of enriching the metallicity than mergers in the early universe.

\item During the merger processes, gas can infall into the central regions of galaxies and form young stellar populations, thereby making the age gradient positive. The fact that the ages of accreted stars in the outer regions by mergers may be relatively old can also contribute to creating positive age gradients. 
\end{enumerate}
We expect that future studies using larger and deeper imaging data will be able to verify our conclusions or increase the statistical significance of our results. Moreover, with these survey data, it will be possible to classify tidal features into various detailed types (tails, shells, streams, diffuse fans, etc.),\footnote{The different types of tidal features may have different origins and lifetimes \citep{Feldmann2008,Tal2009,Mancillas2019}.}  which enables more advanced studies of topics similar to what we have covered here.
\\

\begin{acknowledgments}
This research was supported by the Korea Astronomy and Space Science Institute under the R\&D program (Project No. 2023-1-830-05), supervised by the Ministry of Science and ICT.
\end{acknowledgments}

\appendix

\begin{figure}
\includegraphics[width=\linewidth]{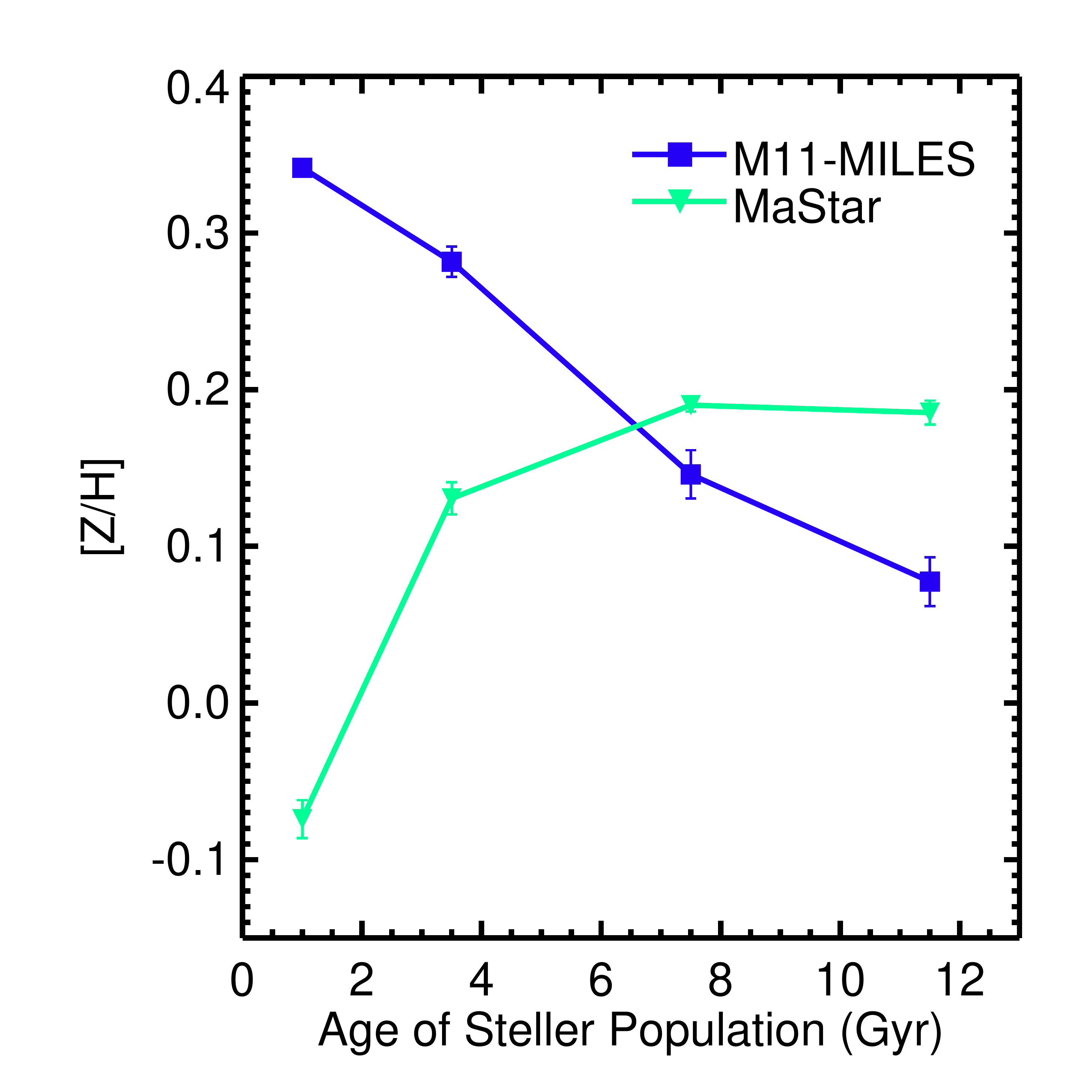}  %width=\linewidth
\centering
\caption{Average metallicity enrichment history for stellar populations of ETGs with $\log(M_\mathrm{star}/M_\odot)\ge10.6$. The four age bins are defined as age/Gyr $<2$, $2\le$ age/Gyr $<5$, $2\le$ age/Gyr $<5$, $5\le$ age/Gyr $<10$, and age/Gyr $\ge10$. Each error bar indicates the standard deviation of the mean values from 1000 bootstrap resamplings. Here, a mass-weighted metallicity is used. We display two cases in which two different stellar population models (M11-MILES and MaStar) are used. See Section \ref{sec:results} for the detailed method for constructing the average metallicity enrichment history.
\label{fig:spcomp}}
\end{figure} 

\begin{figure*}
\includegraphics[scale=0.30]{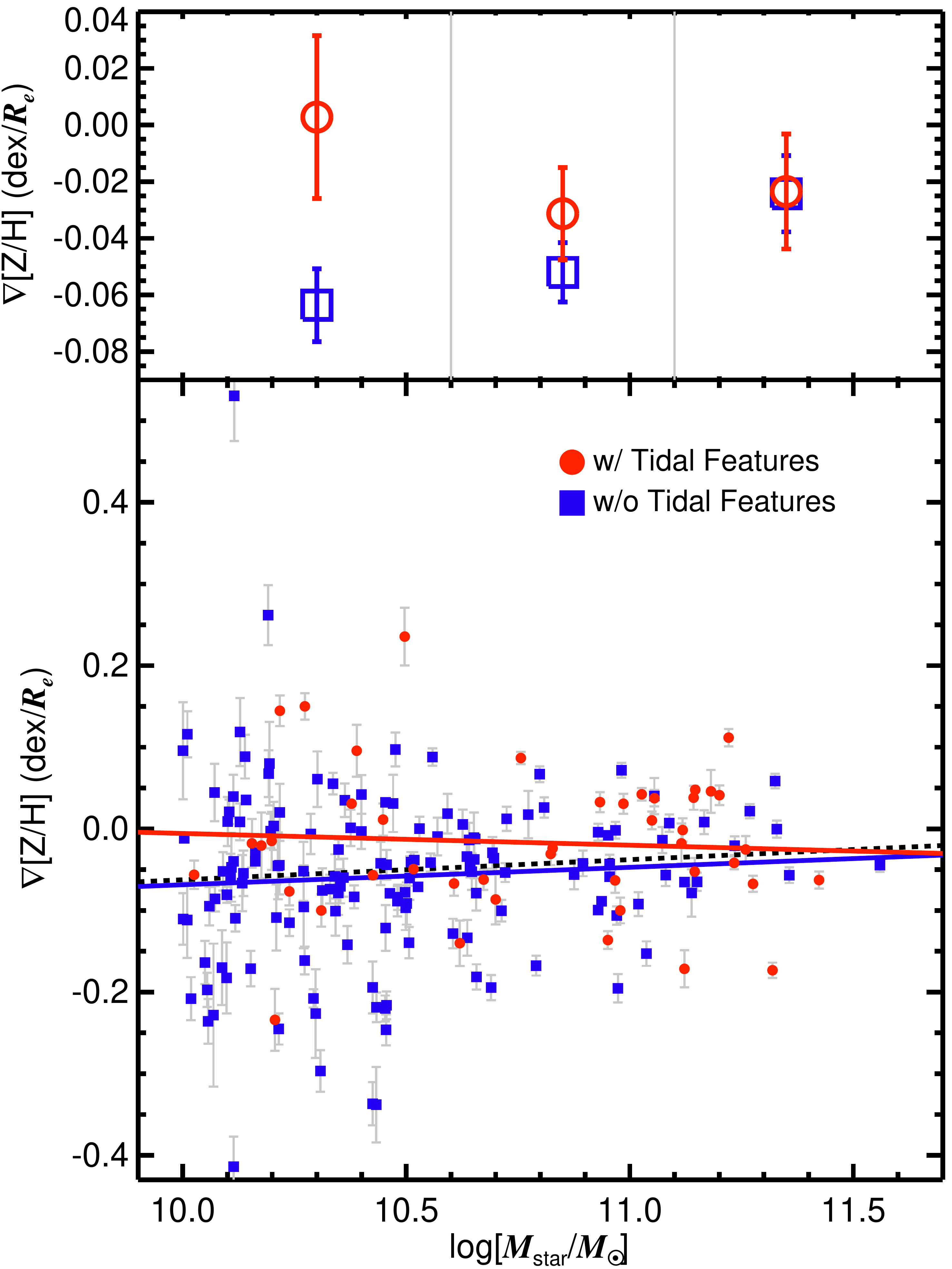}\includegraphics[scale=0.30]{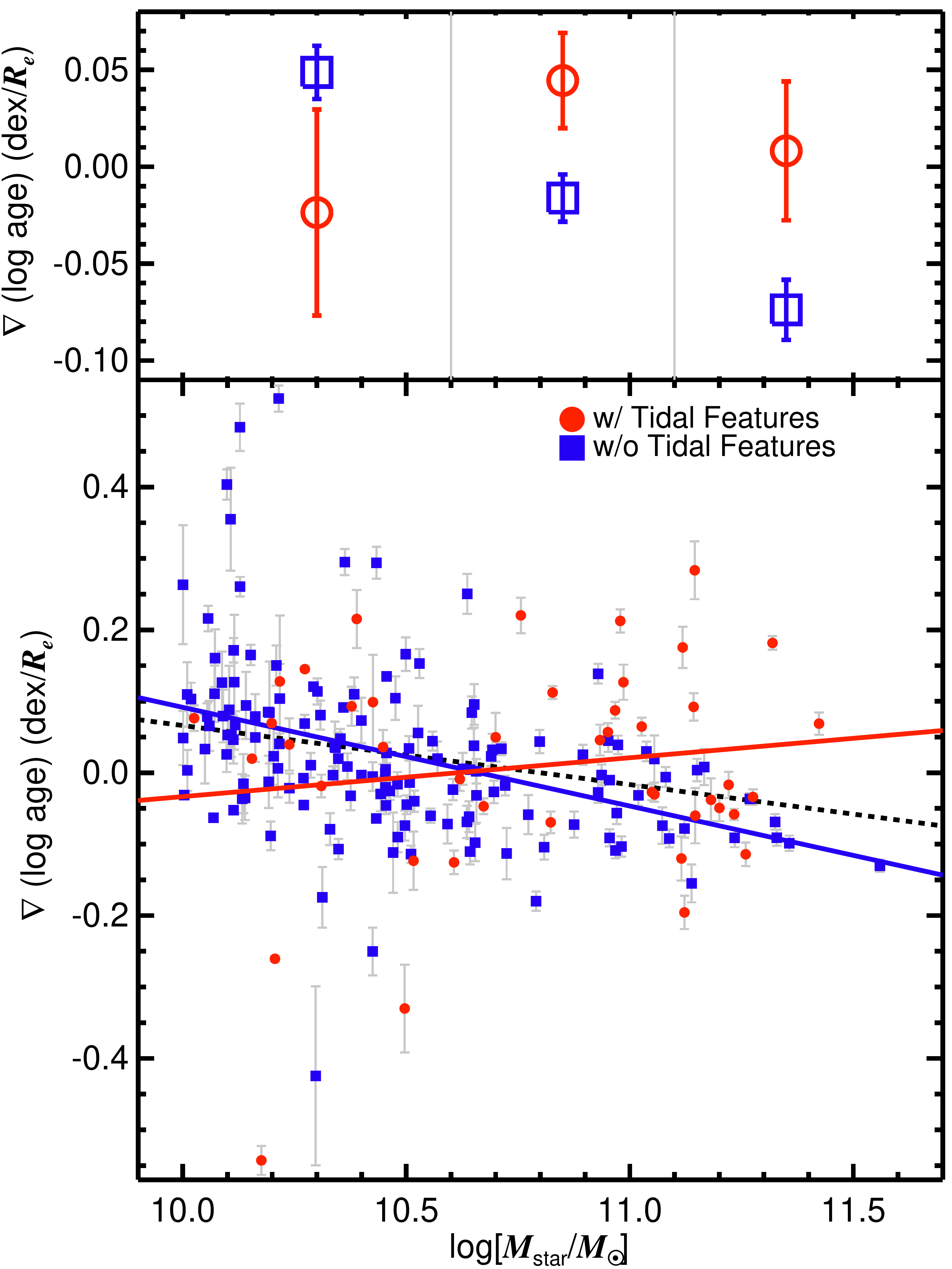}  %width=\linewidth
\centering
\caption{Metallicity gradients (left panel) and age gradients (right panel) of ETGs as a function of galaxy stellar mass. In this figure, the MaStar models are used instead of the M11-MILES models. Other descriptions of the figure are identical to those in Figure \ref{fig:gr}.
\label{fig:gr_ma}}
\end{figure*} 

\begin{figure*}
\includegraphics[scale=0.30]{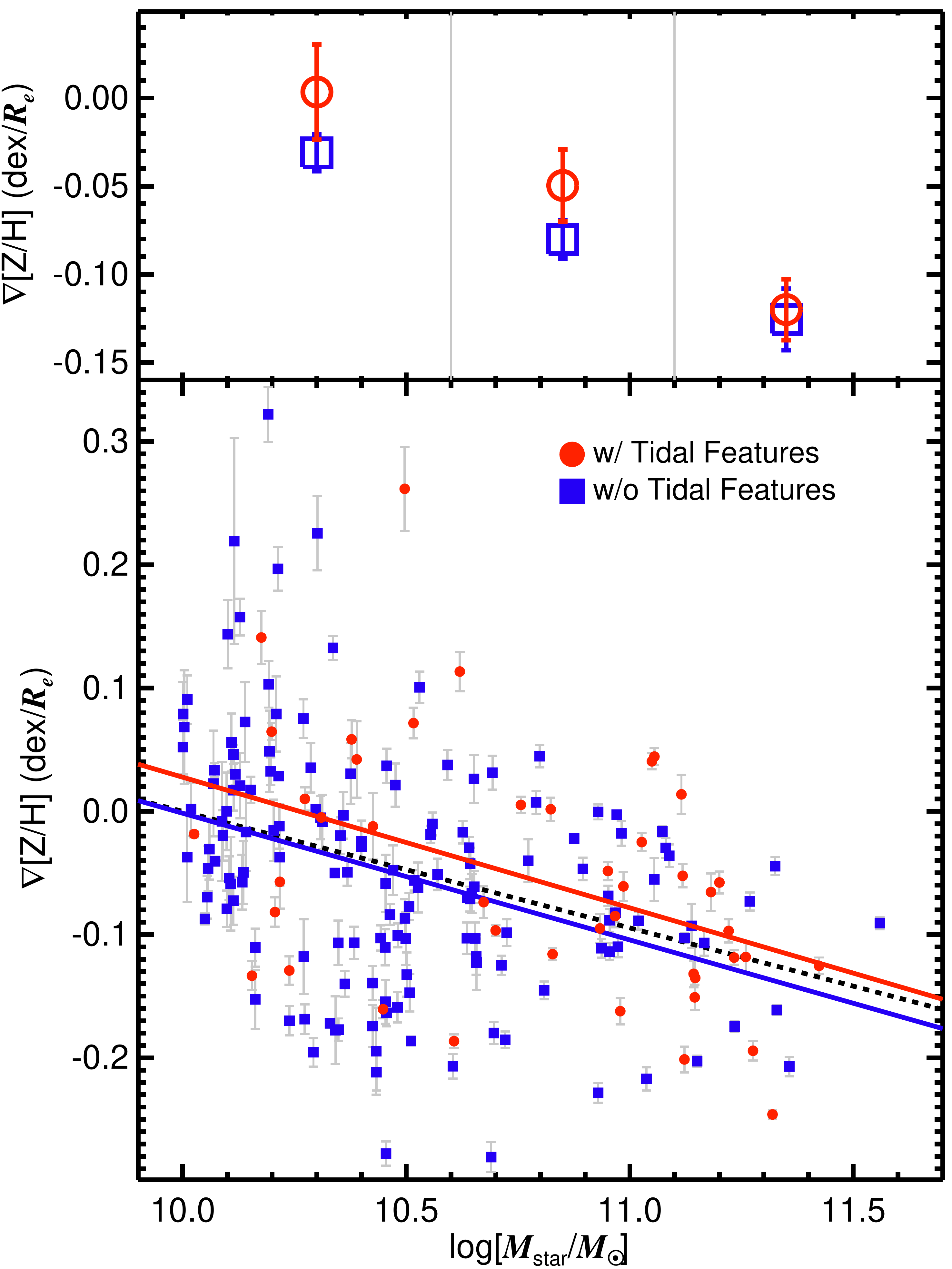}\includegraphics[scale=0.30]{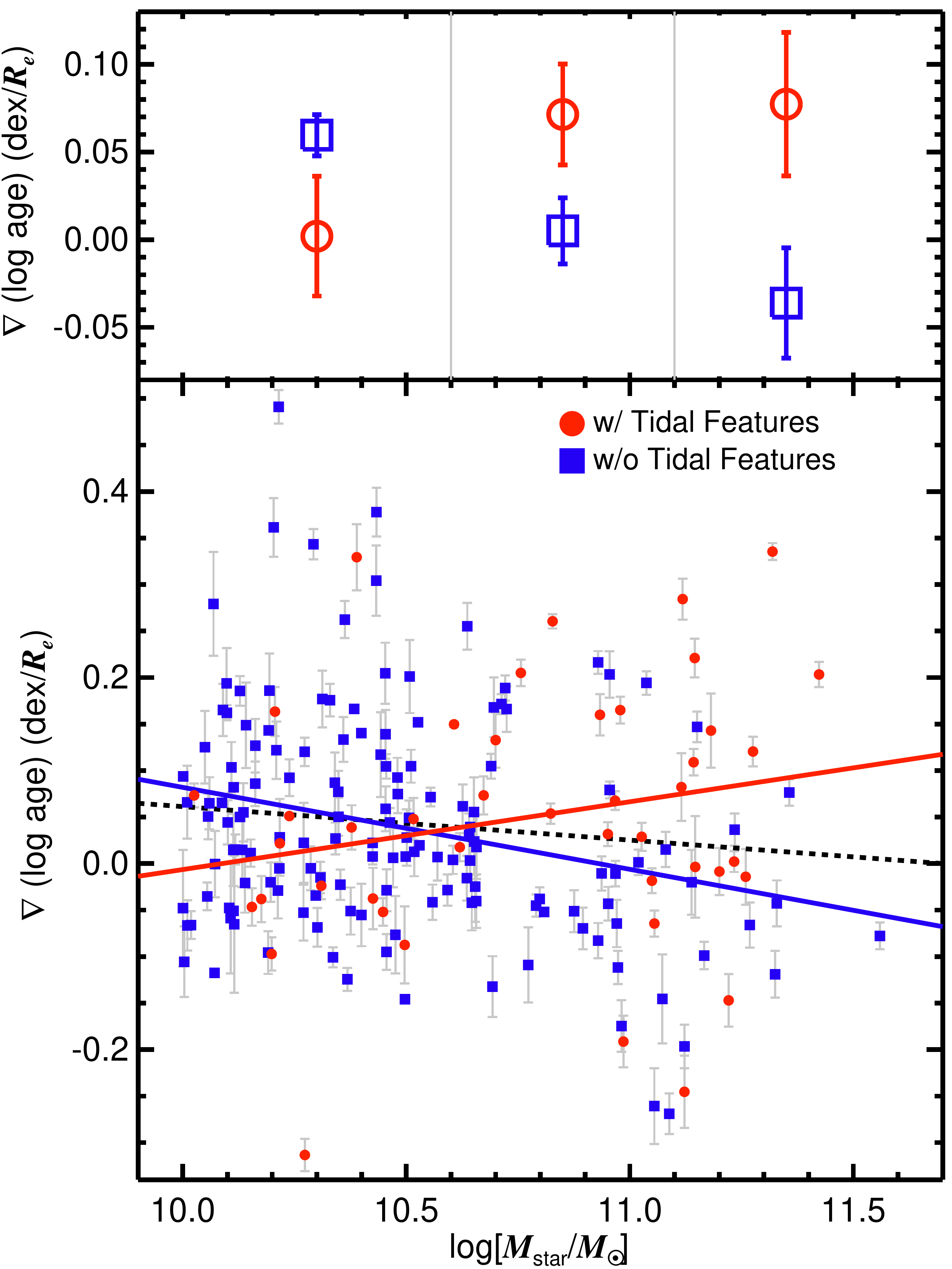}  %width=\linewidth
\centering
\caption{Metallicity gradients (left panel) and age gradients (right panel) of ETGs as a function of galaxy stellar mass. In this figure, light-weighted metallicities and ages (derived from the M11-MILES models) are used instead of the mass-weighted ones. Other descriptions of the figure are identical to those in Figure \ref{fig:gr}.
\label{fig:gr_l}}
\end{figure*} 

\section{Stellar Population Gradients from MaStar Models}\label{Appendix_A}

Here, we provide Figure \ref{fig:spcomp}, which displays the average metallicity enrichment histories for stellar populations of ETGs with $\log(M_\mathrm{star}/M_\odot)\ge10.6$ that are derived from the two different models (M11-MILES and MaStar). 

We also present results for $\nabla[Z/H]$ and $\nabla(\log\mathrm{age})$ derived from the MaStar models in Figure \ref{fig:gr_ma}. As in the case for M11-MILES, ETGs have negative $\nabla[Z/H]$ on average. However, the mass dependence on $\nabla[Z/H]$ for all ETGs is weak and even reversed (the slope of the relation is $0.025\pm0.019$) when compared with that from the M11-MILES models (the slope is $-0.109\pm0.027$). Similar to $\nabla[Z/H]$ from M11-MILES, ETGs with tidal features have shallower $\nabla[Z/H]$ on average than ETGs without tidal features, but only at $\log(M_\mathrm{star}/M_\odot)<11.1$. Conducting the random resamplings, we found that the statistical significance is $98.5\%$ for the fact that ETGs with tidal features at $\log(M_\mathrm{star}/M_\odot)<11.1$ have higher $\nabla[Z/H]$ on average than ETGs without tidal features for a given mass.

The results for $\nabla(\log\mathrm{age})$ derived from the MaStar models shown in Figure \ref{fig:gr_ma} are consistent with those from the M11-MILES models in many respects. As in the results of $\nabla(\log\mathrm{age})$ from M11-MILES, ETGs have a slightly positive or flat $\nabla(\log\mathrm{age})$, except for ETGs without tidal features at $\log(M_\mathrm{star}/M_\odot)\ge11.1$. There is a mass dependence on $\nabla(\log\mathrm{age})$ for all ETGs in the sense that more massive ETGs have lower $\nabla(\log\mathrm{age})$ (the slope of the relation is $-0.083\pm0.023$). This mass dependence is also found on $\nabla(\log\mathrm{age})$ derived from the M11-MILES models (the slope is $-0.032\pm0.024$), but it is more significant in $\nabla(\log\mathrm{age})$ from the MaStar models. At the mass range of $\log(M_\mathrm{star}/M_\odot)\gtrsim10.6$, ETGs with tidal features have more positive $\nabla(\log\mathrm{age})$ on average than ETGs without tidal features, which is the trend that was also found on $\nabla(\log\mathrm{age})$ derived from the M11-MILES models. The statistical significance is $99.74\%$ ($\sim3.0\sigma$) for the fact that ETGs with tidal features at $\log(M_\mathrm{star}/M_\odot)\ge10.6$ have higher $\nabla(\log\mathrm{age})$ on average than the counterparts without tidal features for a given mass.

In conclusion, even though the metal-enrichment history from the MaStar models is opposite to the general cosmic metal-enrichment history (Figure \ref{fig:spcomp} and Section \ref{sec:rspp}), the trends in stellar population gradients derived from the MaStar models are roughly consistent with those from the M11-MILES models, especially in age gradients.
\\

\section{Stellar Population Gradients from Light-weighted Parameters}\label{Appendix_B}

Here, we provide results for $\nabla[Z/H]$ and $\nabla(\log\mathrm{age})$ from light-weighted metallicities and ages (derived from the M11-MILES models) in Figure \ref{fig:gr_l}. We note that the trends in $\nabla[Z/H]$ and $\nabla(\log\mathrm{age})$ from light-weighted stellar population parameters are quite similar to those from the mass-weighted ones described in Section \ref{sec:results} in almost all respects, except that $\nabla[Z/H]$ differences between ETGs with and without tidal features are reduced in the case of light-weighted metallicities (accordingly, the statistical significance is also decreased).
\\

\end{document}